\documentclass{revtex4-1}

\usepackage{amsmath}
\usepackage{amsfonts}
\usepackage{mathrsfs}
\usepackage{amsthm}
\usepackage{graphicx}
\usepackage{subfigure}
\usepackage{amssymb}
\usepackage{hyperref}
\usepackage{url}
\usepackage[abs]{overpic}
\usepackage{epstopdf}
\usepackage[usenames,dvipsnames]{xcolor}
\usepackage{soul}
\usepackage{tikz}

\usepackage{grfext}
\usepackage{grffile}

\PrependGraphicsExtensions*{.png,.PNG}

\def\r{\textbf{r}}

\def\v{\textbf{v}}
\def\U{\textbf{U}}
\def\V{\textbf{V}}
\def\vr{\textbf{v}_{\text{R}}}
\def\va{\textbf{v}_\alpha}
\def\vb{\textbf{v}_\beta}
\def\oma{\pmb{\omega}_\alpha}
\def\x{\textbf{x}}
\def\y{\textbf{y}}
\def\z{\textbf{z}}
\def\gami{\pmb{\gamma}}
\def\gamia{\pmb{\gamma}_\alpha}
\def\gamib{\pmb{\gamma}_\beta}
\def\Pia{\pmb{\Pi}_\alpha}
\def\Pib{\pmb{\Pi}_\beta}
\def\pin{\text{pin\c con}\;}
\def\pins{\text{pin\c cons}\;}
\def\xa{\textbf{x}_\alpha}
\def\xb{\textbf{x}_\beta}
\def\FF{\textbf{F}}
\def\FFa{\textbf{F}_\alpha}
\def\FFb{\textbf{F}_\beta}
\def\rab{\textbf{r}_{\alpha\beta}}

\def\phiab{\phi_{\alpha\beta}}
\def\phia{\phi_{\alpha}}

\newcommand\grad{\mbox{${\bf \nabla}$}}

\newcommand{\beaa}{\begin{eqnarray*}}
\newcommand{\eeaa}{\end{eqnarray*}}

\newcommand{\ovl}[1]{\overline{#1}^\ell}
\newcommand{\ove}[1]{\overline{#1}^\eta}

\nobreak

\begin{document}

\title{{A model of interacting Navier-Stokes singularities}}
\author{
H. Faller$^{1}$, L. Fery$^{1,2}$, D. Geneste$^{1}$ and B. Dubrulle$^{1}$}

\address{$^{1}$ SPEC, CNRS UMR 3680, CEA, Universit\'e Paris-Saclay, 91190 Gif sur Yvette, France\\
$^{2}$ Department of Physics, Ecole Normale Supérieure de Lyon, 69364, Lyon, France}

\date{\today}


\email{hugues.faller@normalesup.org}

\begin{abstract}
We introduce a model of interacting singularities of Navier-Stokes, named \pins. They follow a non-equilibrium dynamics, obtained by the condition that the velocity field around these singularities obeys locally Navier-Stokes equations. This model can be seen as a generalization of the vorton model of Novikov\cite{Novikov83}, that was derived for the Euler equations. When immersed in a regular field, the \pins are further 
transported and sheared by the regular field, while applying a stress onto the regular field, that becomes dominant at a scale that is smaller than the Kolmogorov length. We apply this model to compute the motion of a pair of \pins. A \pin dipole is intrinsically repelling and the \pins generically run away from each other in the early stage of their interaction.  At late time, the dissipation takes over, and the dipole dies over a viscous time scale. In the presence of a stochastic forcing, the dipole tends to orientate itself so that its components are perpendicular to their separation, and it can then follow during a transient time 
 a near out-of-equilibrium state, with forcing balancing dissipation.\
 
 In the general case where the \pins have arbitrary intensity and orientation, we observe three generic dynamics in the early stage: one collapse with infinite dissipation, and two 
 expansion modes, the dipolar anti-aligned runaway and an anisotropic aligned runaway. The collapse of a pair of \pins follows several characteristics of the reconnection between two vortex rings,
 including the scaling of the distance between the two components, following Leray \cite{Leray34} scaling $\sqrt{t_c-t}$.
\end{abstract}

\maketitle

\section{Introduction}
Snapshots of dissipation or enstrophy in turbulent fluids show us that small scales are intermittent, localized and irregular. Mathematical theorems constrain the degree of
irregularity of such 
structures that are genuine singularities of the incompressible Navier-Stokes provided their spatial $\mathbb{L}^3$- norm is unbounded (for a review of various regularity criteria, see \cite{Gibbon16}). On the other hand, dissipation laws of turbulent flows suggest that they may be at most H\"older continuous with $h<1/3$ \cite{C94} and of diverging vorticity in the inviscid limit. This observation has motivated several theoretical construction of turbulent Navier-Stokes small scale structures or weak solutions of Euler equations, using singular or quasi-singular entities based e.g. on atomic like structures \cite{Hicks1899}, Beltrami flows \cite{Constantin88}, Mikado flows \cite{Buckmaster18}, spirals \cite{Lundgren1982,Gilbert93} or vortex filaments \cite{chorin1991}. 


These constructions have fueled a long-standing analytical framework of turbulence, allowing the modeling of proliferating and numerically greedy small scales by
a countable (and hopefully numerically reasonable) number of degrees of freedom, provided by characteristics of the basic entities.\

A good example of the possibilities offered by such a singular decomposition is provided by the 3D vorton description of Novikov \cite{Novikov83}. In this model, the vorticity field is decomposed into $N$ discretized
singularities infinitely localized (via a $\delta$ function) at points $r_\alpha$, ($\alpha=1...N$), each characterized by a vector $\gamia$ providing the intensity and the axis of rotation of
motions around such singularities. The singularities are not fixed, but move under the action of the velocity field and velocity strain induced by the other singularities, so as to respect conservation of circulation. Around the singularity, the velocity field is not of divergence free, so that the vortons are akin to hydrodynamical monopoles interacting at long-range through a potential decaying
like $1/r^2$. The model was adapted to enable numerical simulations of interacting vorticity rings or filaments by considering a divergence-free generalization of the vortons \cite{Aksman85}. Quite remarkably, the Vorton model results in vortex reconnection, even though no viscosity is introduced in the numerical scheme \cite{Alkemade93}. Whether the effective viscosity is due to intense vortex stretching \cite{Pedrizzetti1992},
or to properties of vortex alignment during reconnection \cite{Alkemade93} is still debated. \

From a mathematical point of view, the vorton model cannot be considered as a fully satisfying description of singularities of Navier-Stokes for two reasons.
First, the vortons do not constitute exact weak solutions 
of the 3D Euler or Navier-Stokes equations \cite{Saffman86,Winckelmans88,Greengard88}, which somehow makes them less attractive than point vortices, that are weak solutions of 2D Euler equations\cite{Saffman86}.
Second, vortons do not respect the scaling invariance 
of Navier-Stokes, which imposes that the velocity field should scale like $1/r$. Indeed, through the Biot-Savart law, we see that a Dirac vortex field induces a velocity scaling like $1/r^2$, where $r$ is the distance to singularity.

Motivated by this observation, we introduce in this paper a modification of the vorton model, that is built upon weak solutions of Navier-Stokes equations, and which respects scale invariance of the Navier-Stokes equations, and allows simple dynamical
description of the evolution of the basic entities, herefater named \pins.\

After useful generalities (Section 2.a), we introduce the \pin model (Section 2.b) and describe their properties in Section 2.c. We introduce the non-equilibrium dynamics of \pins in Section 2.d and 2.e. We then solve the equations in Section 3, starting with the special case of a dipole in Section 3.b and 3.c, and concluding with the general case in Section 3.d.
A discussion follows in Section 4.

\section{Pin\c con model}
\subsection{Generalities and ideas behind \pin model}
Consider a velocity field $\U$ obeying the Navier-Stokes equations. Then it is well known that the coarse-grained field $\ovl{\U}$ obeys the Navier-Stokes equations forced by the "turbulent force" 
due to the Reynolds stress $\grad\cdot\left( \ovl{\U}\ovl{\U}- \ovl{\U \U}\right)$. Numerical and experimental observations also show that as $\ell\to 0$, this turbulent force becomes more and more intermittent, made of isolated patches of finite values, in a sea of zero values. The size of the isolated patches shrinks with decaying $\ell$. In our experiment, we have observed that such patches
do persist even when $\ell$ is of the order of the Kolmogorov scale $\eta_K$ , and have correlated such patches with the existence of nonzero energy local energy transfers at such location \cite{geneste2021}.
This means that numerical simulations of Navier-Stokes must have a resolution much smaller than $\eta_K$ in order to fully resolve not only velocity gradients \cite{Yeung18} but also 
local energy transfers and dissipation \cite{Dubrulle19}. The numerical price to pay is high, especially at large Reynolds number, and a lot of computing time is wasted in the tracking of
increasingly thinner regions of space.\

To avoid this, a natural idea is to split the fluid in two component: one, continuous, representing the coarse-grained fluid $\ovl{\U}$, for a scale $\ell=\ell_c$ to be determined later, and one, discrete, representing the isolated patches of unresolved fluid, that are fed by the turbulent force and carry and dissipate the corresponding energy with a dynamics to be determined later. In this view, the small scales
must therefore be represented by modes, that are representative of the small scale behaviour of Navier-Stokes. Given that we want to be able to describe the whole range of scales $\ell<\ell_c$,
it is natural to consider self-similar solutions of Navier-Stokes, i.e. solutions that are invariant by the (Leray) rescaling $\U(\x,\,t) \to \lambda^{-1}\U\left(\x/\lambda,\,t/\lambda^{2}\right)$ for any $\lambda$ \cite{Leray34}. Moreover, to be able to describe the small scales by modes dynamics, we wish to consider self-similar solutions that do not explicitly depend on time, so that all the dynamics will be contained in the time variation of the modes parameters. Corresponding solutions then obey 
\begin{equation}
\forall \lambda\neq 0,\U(\x) = \lambda^{-1}\U\left(\x/\lambda \right),
\label{homogeneous}
\end{equation}
corresponding to homogeneous Navier-Stokes solutions of degree -1.\

As shown by Sverak \cite{Sverak2011}, Theorem 1, the only non-trivial solutions that are smooth in in $R\backslash \{0\} $ are axisymmetric, and correspond to Landau solutions, described in Section
\ref{SecPincon}, equation \ref{LandauSolution}. These solutions 
 obey the stationary Navier-Stokes equations everywhere except at the origin. Specifically, we have in some distributional sense:
\begin{eqnarray}
\nabla\cdot \U&=&0,\nonumber\\
{(\U\cdot\grad)\U} +\frac{\grad {p}}{\rho}- \nu\Delta\U
&=&\nu^2 \delta(\x){\bf F},
\label{equaforgyron}
\end{eqnarray}
where ${\bf F}$ is a vector magnitude $F$ and given orientation ${\bf e}$, providing the axis of symmetry.\

We thus do not have much choice in the building of our small scales modes. Here is how we build them, using Landau solutions.

\subsection{Definition of \pin}
\label{SecPincon}

\begin{figure}
\includegraphics[width=0.5\textwidth]{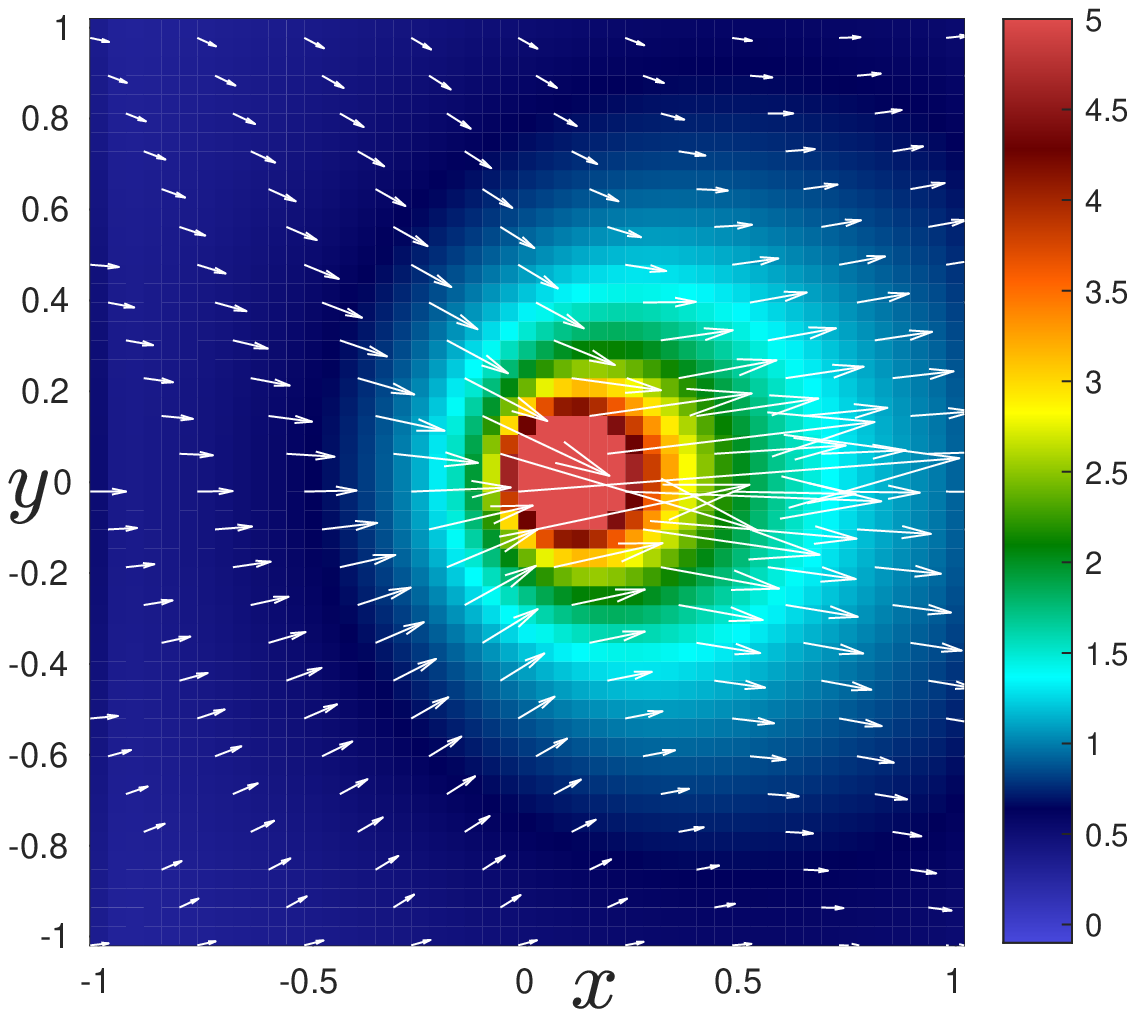}
\put(-110,158){\bf(a)}
\put(-43,160){$\textbf{v}_\alpha \cdot \textbf{e}_z$}
\includegraphics[width=0.5\textwidth]{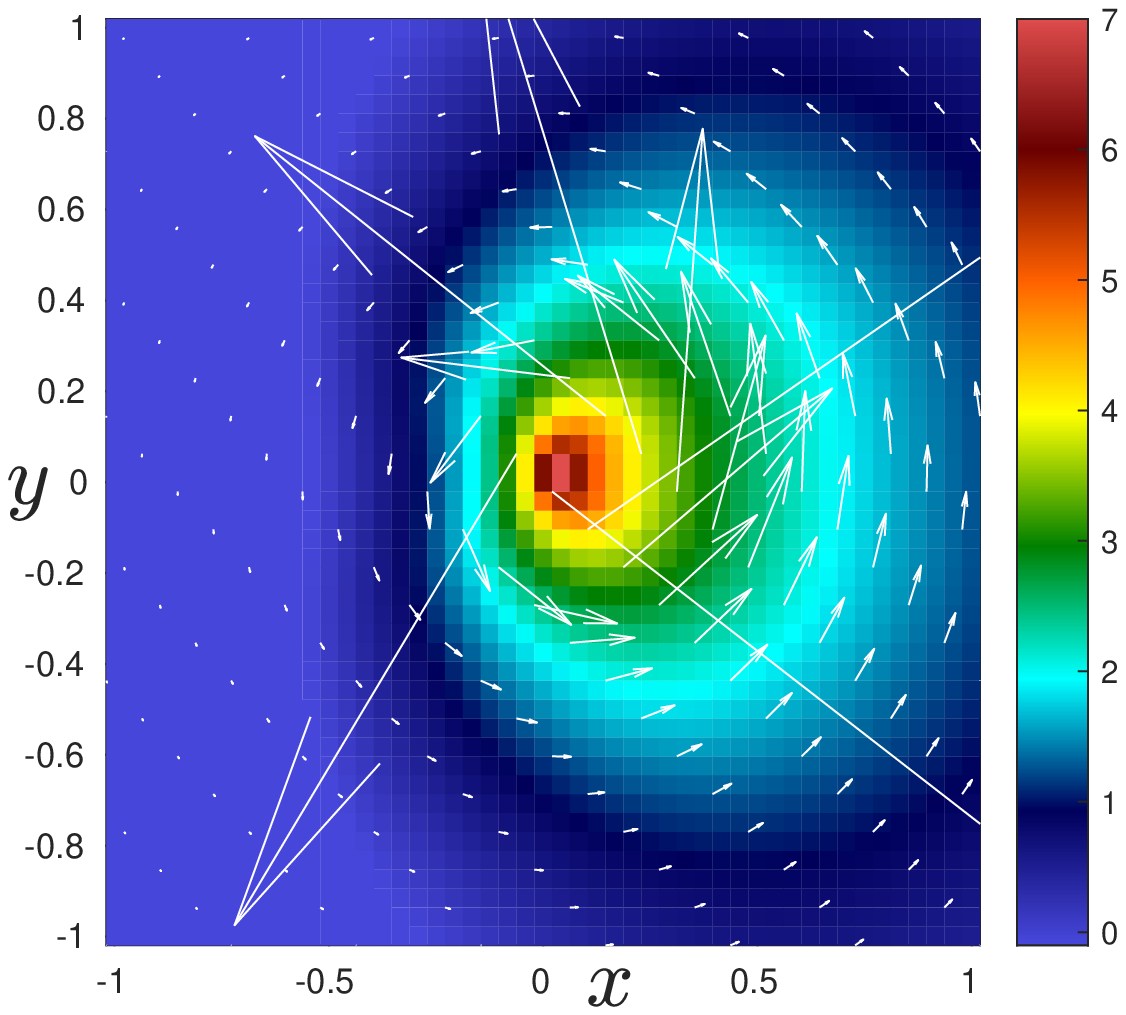}
\put(-110,158){\bf(b)}
\put(-45,160){$\log(\|\omega\|)$}
\caption{(a) Velocity and (b) vorticity field around a \pin. The white arrows provide the velocity and vorticity in the plane generated by $\x$ and $\gamia$. The colour codes the out of plane velocity (a) and the logarithm in base 10 of the enstrophy (b).}
\label{fig:pincon}
\end{figure}

\begin{figure}
\includegraphics[angle=-90,width=0.5\textwidth]{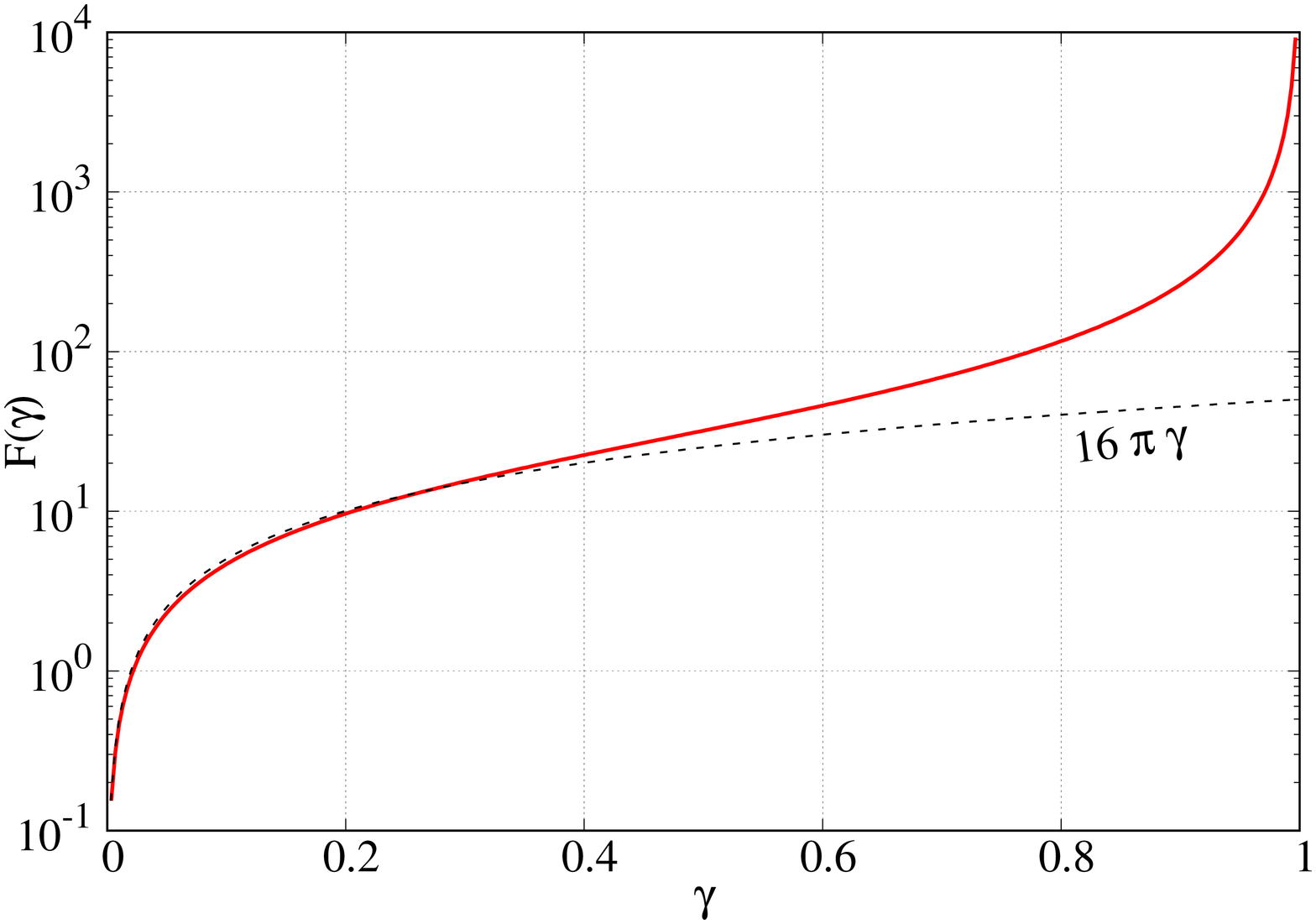}
\put(-120,-110){\bf(a)}
\includegraphics[angle=-90,width=0.5\textwidth]{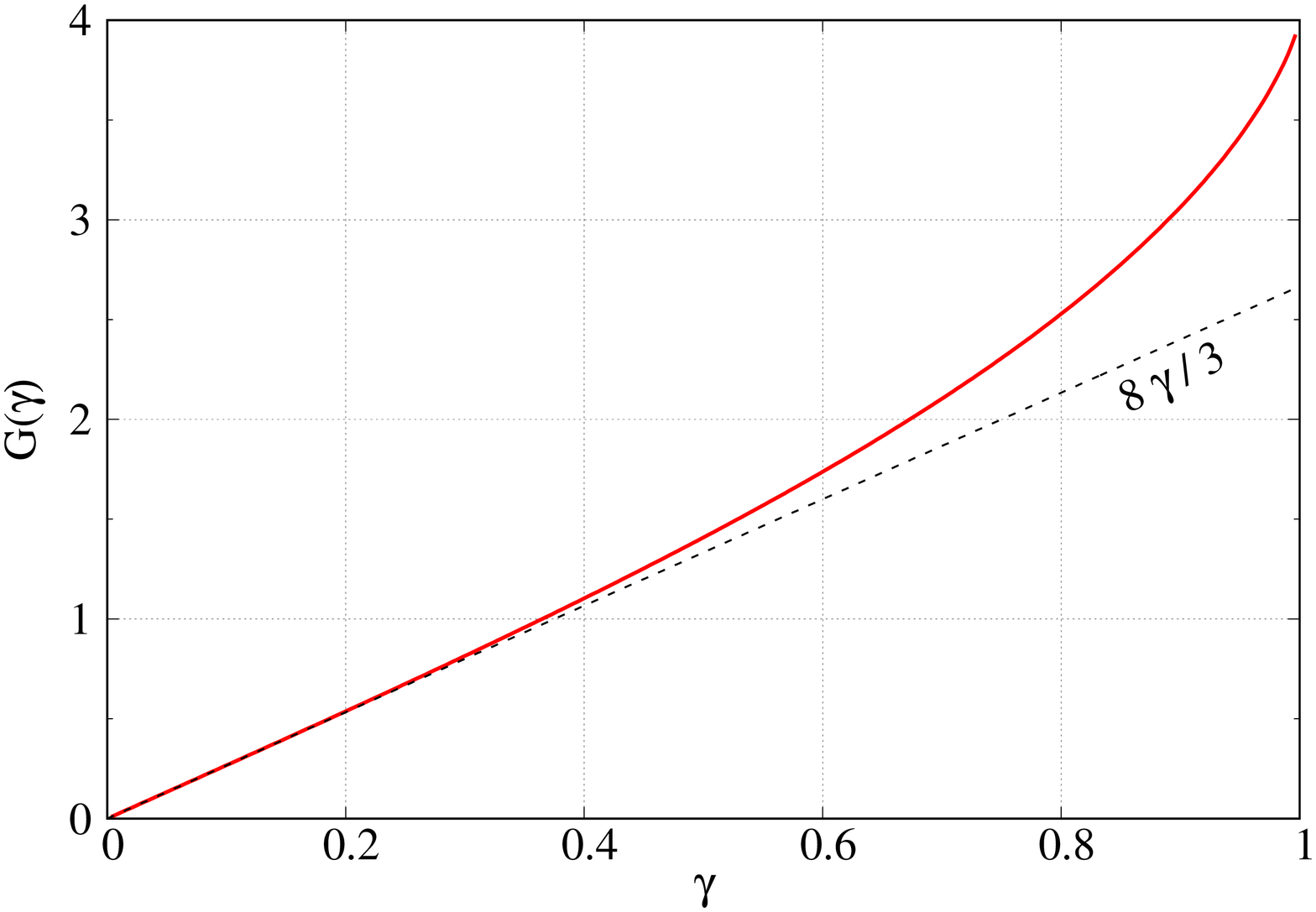}
\put(-120,-110){\bf(b)}
\caption{Parameters of a \pin as a function of its intensity $\gamma$. (a) Intensity of the force produced by the \pin at its location. The black dashed line has equation $y=16\pi \gamma$; (b) Generalized momentum of a $\pin$. The black dashed line has equation $y=8 \gamma/3$. }
\label{fig:prop}
\end{figure}

We introduce the \pins as individual entities labeled by $\alpha$, characterized by their position $\x_\alpha(t)$, and a vector $\gamia(t)$, with $\|\gamia\|<1$, that 
 produce locally an axisymmetric velocity field around the axis of direction $\gamia$ given by $(p_\alpha,\va)(\x)\equiv (\nu^2 p(\x-\xa,\gamia),\nu \U(\x-\xa,\gamia))$ with $\U$ and $p$ given by:
\begin{eqnarray}
\U(\x-\xa,\gamia)&=&\frac{2}{\phi_\alpha}\bigg(\gamia-\frac{\x-\xa}{\|\x-\xa\|}\bigg)+2(1-\gamia^2)\frac{\x-\xa}{\phi^2_\alpha},
\label{veloalpha}\\
p(\x-\xa,\gamia)&=&-\frac{4}{\|\x-\xa\|\phi_\alpha}+4\frac{1-\gamia^2}{\phi^2_\alpha}.
\label{LandauSolution}
\end{eqnarray}
where $\phi(\x,\gami)=\|\x\|-\gami\cdot\x$ and $\phi_\alpha=\phi(\x-\x_\alpha,\gamia)$ and $p_\alpha$ is the associated pressure. A few useful properties of $\phi$ are put in Appendix. In particular, the velocity field given by Eq. (\eqref{veloalpha})
is homogeneous of degree -1 around $\xa$, and axisymmetric around the direction of $\gami$. 
Plots of velocity and vorticity around a \pin are displayed in figure \ref{fig:pincon}. Close to the singularity, there is a neck pinch of the velocity streamlines, hence their name 
\pin.
 As first shown by Landau \cite{Landau44} (see also \cite{Squire52,Batchelor2000,tian1998,Cannone2004,Sverak2011}), the velocity fields $\v_\alpha$ are solutions of (\eqref{equaforgyron}) with 
\begin{eqnarray}
{\bf F}&=&F(\gamma)\frac{\gamia}{\gamma},\nonumber\\
F(\gamma)&=&4\pi\left[\frac{4}{\gamma}-\frac{2}{\gamma^2}\ln\bigg(\frac{1+\gamma}{1-\gamma}\bigg)+\frac{16}{3}\frac{\gamma}{1-\gamma^2}\right],\\
\gamma&=&\|\gamia\|.\nonumber
\label{forcepincon}
\end{eqnarray}
We refer the reader to \cite{Cannone2004} for a rigorous derivation of such result. The function $F(\gamma)$ is shown in Fig. \ref{fig:prop}-a. It starts from $0$ at $\gamma=0$, with a linear behaviour $F(\gamma)=16\pi\gamma$ near the origin, and diverges at $\gamma=1$.

\subsection{Properties of \pins}

\subsubsection{Scaling under coarse-graining}
The velocity field and all its derivatives diverge at the location of the \pin so they are undefined as such point. We may however 
 study its behavior near the origin by introducing a suitable test function $\psi$ that is spherically symmetric around $x=0$, positive of unit integral, $C^\infty$ and that decays fast at infinity, and considering the regularizations
\begin{equation}
\ovl{\va}(\x)=\int \psi\left(\frac{\x-\y}{\ell}\right) \va(\y) \frac{\mathrm d\y}{\ell^3},\nonumber\\
\label{regul}
\end{equation}
where $\ell$ is a small parameters. In the limit $\ell\to 0$, the function $\psi\left(\frac{\x}{\ell}\right) $ is peaked around the origin so that, as long as $\x$ is far from $\xa$, we can estimate:
$ \ovl{\va}(\x)\approx \va(\x)$. Consider now the situation where $\x=\xa$. We have then:
\begin{equation}
\ovl{\va}(\xa)= \int \psi\left(\frac{\xa-\y}{\ell} \right) \U(\y-\xa,\gamia) \frac{\mathrm d\y}{\ell^3}.\nonumber\\
\label{regul1}
\end{equation}
Applying finally the change of variable $\y-\xa=\ell \z$, using homogeneity properties of $\U$ and spherical symmetry of $\psi$ we have :
\begin{eqnarray}
\ovl{\va}(\xa)&=&\frac{1}{\ell} \int \psi\left(\z \right) \U(\z,\gamia) {\mathrm d\z},\nonumber\\
&=& \frac{C_\psi}{\ell}\langle \va \rangle_{{\cal B}_1}\underset{\ell \rightarrow 0}{=}\mathcal{O}(\ell^{-1}),
\label{regul2}
\end{eqnarray}
where $C_\psi= 4\pi \int r\psi(r) \mathrm{d}r$ and $\langle \va \rangle_{{\cal B}_1}$ is the average over the sphere of radius unity. Via Euler theorem, $\nabla \va$ is homogeneous of order -2. By the same 
reasoning, we then find that $\ovl{\nabla \va}(\xa)\underset{\ell \rightarrow 0}{=}~\mathcal O(\ell^{-2})$. We cannot apply the same reasoning to $\nabla^2 \va$ because the integral $\int \psi(r) \mathrm{d}r/r$ does not necessarily converge
at the origin. However, we have the property that $\ovl{(\va\cdot\grad)\va} +\frac{\grad \ovl{p_\alpha}}{\rho}- \nu\Delta\ovl\va =\frac{\nu^2 }{\ell^3}\psi\left(\frac{\x-\xa}{\ell} \right){\bf F}$ which is 
$\mathcal O(\ell^{-3})$.

\subsubsection{Potential vector, vorticity and helicity}
Using vector calculus identities, we can check that the velocity field around a \pin derives from the vector potential:
\begin{eqnarray}
\textbf{A}_\alpha(\x)&=&2\nu(\x-\xa)\times \nabla \ln (\phia),\nonumber\\
&=&2\nu\frac{\gamia\times (\x-\xa)}{\phia},
\label{Aalpha}
\end{eqnarray}

We can formally define the vorticity field produced locally around a \pin by taking the curl of $\v_\alpha$. The vorticity is parallel to the potential vector and reads:
 \begin{equation}
\oma(\x)=4\nu(1-\gamma^2)\frac{\gamia\times (\x-\xa)}{\phia^3},
\label{omalpha}
\end{equation}
From this, we see that the velocity field produced by a \pin is of zero helicity.

\subsubsection{Generalized Momentum and coarse-grained vorticity}
We define a generalized momentum $\Pia$ for the \pin as 
average of the velocity field over a sphere of unit radius (see appendix for its computation):
\begin{eqnarray}
\Pia\equiv \langle \va \rangle_{{\cal B}_1}&=& \nu G(\gamma_\alpha) \frac{\gamia}{ \gamma_\alpha},\nonumber\\
G(\gamma)&=&\frac{2}{\gamma^2} \left[2\gamma-(1-\gamma^2)\ln\bigg(\frac{1+\gamma}{1-\gamma}\bigg)\right].
\label{GenMom}
\end{eqnarray}
By definition, $\Pia$ provides an estimate of the coarse-grained velocity field at the \pin position, via $\ovl{\va}(\xa)=~C_\psi\Pia/\ell$. Note that $\Pia$ points in the direction of $\gamia$. For $0\le\gamma <1$, $G(\gamma)$ varies smoothly from $0$ to $4$, starting from a linear behavior $G(\gamma)=8\gamma/3$ at the origin and ending with a vertical tangent at $\gamma=1$ 
 (see Fig. \ref{fig:prop}-b). Therefore the function $G(\gamma)$ is bijective, and there is a one-to-one correspondance between $G$ and $\gamma$ and $\pmb{\Pi}$ and $\gami$.\
 
 Due to the axisymmetry, we see that $\ovl{\omega}(\xa)=0$, so that the coarse-grained \pin is vorticity free near its location.

\subsection{Interaction of \pin with a regular field}
There are several reasons why the notion of a "single" \pin does not make a physical sense:
\par i) the \pin is dissipative, and requires a force to maintain it; a surrounding fluid can provide the necessary forcing;
\par ii) the \pin lives in a infinite universe and does not fulfill the boundary conditions of a realistic system. To be able to use a \pin in confined systems, we need to add an external velocity field that will take care of the boundary conditions; 
\par iii) if we accept that the \pin describes the very intermittent part of the energy transfer that cannot be resolved, we must also accept the possibility of coexistence and interaction of several \pins. If we assume that the \pins are distincts and that they are regular everywhere except at their position, this amounts again to consider the interaction of a \pin with a regular field.\

Given the nonlinear nature of the Navier-Stokes equations, we cannot superpose two solutions of Navier-Stokes equations (a \pin and a regular field) to get a solution of Navier-Stokes equations. Instead, we will now assume that the \pin has a dynamics and find such dynamics by imposing that the superposition of the \pin and of the regular field obeys locally the Navier-Stokes equations.
Specifically, we consider a solution of the shape $\va=\U(\x-\xa(t), \gamia(t))$ and ${\bf F}=F(\gamma(t)) {\bf e(t)}$, where $F$ is a prescribed function, and $\xa$, $\gami= \gamma \bf{e}$, two vectors that parametrize 
the field $\va$ as a function of $t$. We then introduce 
$\v=\vr+\va$, where $\vr$ is a velocity field that is regular at the origin, and we impose that $\v$ is a solution of Navier-Stokes locally around the singularity at $\xa$, i.e. that $\v$ is a solution of 
\begin{equation}
\partial_t \ovl{\v}(\xa) + \ovl{(\v\cdot\grad)\v}(\xa)+\frac{\ovl{\grad p}(\xa)}{\rho}-\nu\ovl{\Delta \v}(\xa)=0.
\label{mNSE}
\end{equation}

 Decomposing the velocity field into its regular and irregular part, we see that Eq. \eqref{mNSE} generates terms of various orders in $\ell$, that scale according to Table 1. Note that since $\vr$ is a regular field, its coarse-graining scales like $O(1)$, so do its derivatives. Furthermore, we introduce the quantity $\tau^\ell= \ovl{\vr}\ovl{\vr}- \ovl{\vr \vr}$ which is the Reynolds stress contribution due to filtering. This term has a different scaling. Indeed, $\nabla \cdot \tau^\ell=~\mathcal{O}(\delta\pmb{v}_\ell^2/\ell)$ \cite{EyinkLN}, where {$\delta \pmb{v}_\ell=~\vr(\x+~\ell)-~\vr~(\x)$}.
 Since $\vr$ is regular, it can be expanded as $\vr(\x+~\ell)-~\vr(\x)=~\ell\grad\vr$, so that $\nabla \cdot \tau^\ell \sim \ell (\grad \vr)^2=O(\ell)$.
 
 Collecting the different term we find that the l.h.s. of Eq. \eqref{mNSE} is the sum of the following orders:
\begin{eqnarray}
&\mathcal{O}(\ell)&:
-\nabla \cdot \tau^\ell\\
 &\mathcal{O}(1)&:
\partial_t \ovl \vr + \grad\cdot \ovl{\vr}\ovl{\vr} + \frac{\grad\ovl{p_\text{R}}}{\rho}
-\nu\Delta\ovl \vr \\
&\mathcal{O}(1/\ell)&:\dot{\gami}\pmb{\grad}_\gamma \ovl\va+\ovl{(\va\cdot\grad)\vr}\\
&\mathcal{O}(1/\ell^2)&:-\dot{\x}_\alpha\grad_{\x}\ovl\va +\ovl{(\vr\cdot\grad)\va}\\
&\mathcal{O}(1/\ell^3)&:\ovl{(\va\cdot\grad)\va} +\frac{\grad \ovl{p_\alpha}}{\rho}- \nu\Delta\ovl\va 
\end{eqnarray}

\begin{table}
\centering
\begin{tabular}{|c|c|c|c|c|c|}
\hline
$\v_\alpha$ & $\dot{\gamia}\partial_\gamma(\v_\alpha)$ & $\dot{\x}_\alpha\partial_x\v_\alpha$& $(\v_\alpha\cdot\grad)\vr$&$(\vr\cdot\grad)\v_\alpha$ &$\grad\cdot \ovl{\vr}\ovl{\vr} $\\
\hline
$1/{\ell}$ & $1/\ell$& $1/{\ell^2}$ &$1/{\ell}$&$1/{\ell^2}$& $1$\\
\hline
\end{tabular}
\caption{Order of the various terms appearing in equation \eqref{mNSE} as a function of the filter length $\ell$ in the limit $\x\to \xa$.}
\end{table}

Cancelling the $\mathcal{O}(1/\ell^2)$ provides a first condition as:
 \begin{equation}
\dot{\x}_\alpha\grad_{\x}\ovl\va =\ovl{(\vr\cdot\grad)\va}.\\
\label{condi1}
\end{equation} 
Due to the regularity of $\vr$, we can write $\ovl{(\vr \cdot\grad)\va}=\vr(\xa) \cdot \ovl{\grad \va}$ for small enough $\ell$. Condition \eqref{condi1} is then satisfied providing:
\begin{equation}
\dot{\x}_\alpha =\vr(\xa).
\label{Mov1}
\end{equation} 
Physically, this means that the singularity point is advected by the regular field surrounding it.\

Cancelling the $\mathcal{O}(1/\ell)$ provides a second condition, as: 
\begin{equation}
\dot{\gami}\ovl{\grad_\gamma\va}=-\ovl{(\va \cdot\grad )\vr}.
\label{condi2}
\end{equation}
Due to the regularity of $\vr$, we can write $(\ovl{\va \cdot\grad)\vr}=(\ovl{\va} \cdot\grad)\vr(\xa)$. We then get the equation:
\begin{equation}
\dot{\gami} \ovl{\grad_\gamma\va}=-(\ovl{\va} \cdot\grad)\vr.
\label{Mov2}
\end{equation}
 Physically, this means that the force axis and its direction are moved around by the shear of the regular field at the location of the singularity. \
 
 Cancelling the $\mathcal{O}(1)$ term provides the condition that $\ovl{\vr}$ is a solution of the Navier-Stokes equation. Indeed, this is the idea behind the two fluid model, to allow for 
 the scales above the coarse-graining to be described by a solution of the Navier-Stokes equation.\
 
 We are then left with the smallest order $O(\ell^2)$ and the highest order term $O(\ell^{-3})$, which cannot be balanced in general. For the system to have a physical solution,
 we then impose a "bootstrap condition", namely that the two terms must be of the same order of magnitude, thereby fixing the coarse-graining scale $\ell_c$ via $\nu^2/\ell_c^3\sim~\ell_c (\grad\vr)^2$. We thus find $\ell_c=(\nu^3/\epsilon_r)^{1/4}$, with $\epsilon_r=\nu(\grad\vr)^2$ is the dissipation of the regular field. Therefore, the coarse-graining scale imposed by
 the bootstrap condition is precisely the Kolmogorov scale $\ell_c=\eta$.
 
 Physically, this condition can be understood as follow: the singularity is dissipative, and to maintain it, one must apply a force. Such force is provided by the regular field, through the term $\nabla\cdot\tau^\ell$, which keeps track of the fraction of
 the velocity field that is sent to the subgrid scale, and which is taken into account by the \pin. Conversely, the \pin applies an extra turbulent stress onto the regular field, that extends around
 it in a ball of radius of the order of the Kolmogorov scale. To keep a precise account of these effects, we thus split the Reynolds stress and the \pin force into a contribution at $\xa$ and a 
 contribution around that location, ans share the contribution among the pin\c con, and the regular field.\
 
 Taking into account the fact that $\ove{\vr}\approx \vr$, $\ove{\va} = C_\psi \Pia/\eta$ and $\dot{\gami} \ove{\grad_\gamma\va}\approx C_\psi \dot \Pia/\eta$, we then get the following system of equations to describe the coupling
 between the \pin and the regular field:
 \par for the regular field:
 \begin{equation}
 \begin{array}{rcl}
  \partial_t \ove{\vr}(\x) &+& (\ove{\vr}\cdot\grad)\ove{\vr}(\x)+\frac{\grad\ove{p_R}}{\rho}(\x)-\nu\Delta\ove \vr(\x)\\
 &=& \nabla \cdot \tau^\eta(\x)-\frac{\nu^2 }{\eta^3}\psi\left(\frac{\x-\xa}{\eta} \right)\FFa;
  \end{array}
  \label{equalargescale}
 \end{equation}
 \par for the pin\c con:
 \begin{eqnarray}
 \dot{\x}_\alpha &=&\ove{\vr}(\xa),\nonumber\\
 \dot \Pia&=&-(\Pia \cdot\grad)\ove{\vr}(\xa)+\frac{\eta}{C_\psi} \nabla \cdot \tau^\eta(\xa)-\frac{\nu^2 \psi(0)}{C_\psi\eta^2}{\FFa}.
 \label{equalsmallscale}
 \end{eqnarray}

 These equations describe a two-fluid approach of turbulence, coupling a coarse-grained field at the Kolmogorov scale, and the \pin.
The velocity fluid is follows coarse-grained Navier-Stokes equations, with the Reynolds stress providing the necessary driving force to create \pins. The latter are entities living below the Kolmogorov scale, that are 
 advected and sheared by the coarse-grained fluid, and exert back a forcing on the coarse-grained field, that results in a dissipation of energy. To conclude our two-fluid model, we 
 must prescribe the interactions between \pins, in a way that is compatible with the Navier-Stokes equations.

\subsection{Interactions of \pins}
An ensemble of $N$ \pins, $\alpha=1...N$ produces a velocity field $\v(x,t)$:
\begin{equation}
\v(x,t)=\sum_{\alpha} \va(x,t),
\label{velocityfield}
\end{equation}

Around a \pin \;\;$\alpha$, the ensemble of other \pins produces a regular field $\vr=~\sum_{\beta\neq \alpha} \vb(\xa)$. Motivated by such observation, we {\sl define} the interactions of the \pins via the  following set of $2N$ differential equations:
\begin{eqnarray}
\dot{\x}_\alpha&=&\sum_{\beta\neq \alpha} \vb(\xa,t),\\
\dot{\pmb{\Pi}}_{\alpha}&=&-(\Pia \cdot \grad_{\xa})\bigg( \sum_{\beta\neq \alpha} \vb(\xa,t)\bigg)-\frac{\nu^2 \psi(0)}{C_\psi\eta^2}\FFa+\frac{E_\alpha}{C_\psi}\chi,
\label{interaction}
\end{eqnarray}
where $E_\alpha=U_\alpha^2$ describes the local energy of the large scale regular field that provides a stochastic forcing $\chi$ via the (random) Reynolds stress contribution. In the sequel, we assume that $\chi$ is isotropic and shortly correlated over a a Kolmogorov time scale, so that $\langle \chi_i(t)\chi_j(t')\rangle=~\delta(t-~t')\delta_{ij}$. Note that the equation \eqref{interaction}
is a definition, that leaves aside many conditions that may have to be satisfied for the model to be an exact representation of the small scales of Navier-Stokes. For example, this model is more likely to be valid as the dilute limit is achieved, so that the \pins are sufficiently apart from each other for them to be considered as ponctual particles. Also, we do not try to make a difference between 
close and distant interactions, while in the former case, diverging velocities and correlations may impede our possibility to consider that the field generated by the external fields is smooth enough so that the approximation $\ovl{(\vr \cdot\grad)\va}=\vr(\xa) \cdot \ovl{\grad \va}$ is valid for small enough $\ell$. Therefore, even if individually each \pin is a weak solution of Navier-Stokes equations,
the collection of N \pin is not an exact weak solution of Navier-Stokes equations.\

In some sense anyway, the equations of motions of the \pins correspond to the equations that are imposed by the structure of the Navier-Stokes equations and the requirement that the local velocity 
field induced by each \pin should obey such equations. Eqs. \eqref{interaction} can therefore be seen as the equivalent of the motion of poles or zeros of partial differential equations that have been 
computed, starting from Kruskal\cite {Kruskal74} for the KdV equations (see \cite{Calogero96,Calogero01} for a review). The motions
are furthermore constrained by the condition that they stay with the unit hypersphere such that $\| \gamia\| < 1$.\

The \pin are characterized by an interaction energy:
\begin{equation}
E= \frac{C_\psi}{\eta} \sum_{\beta\neq \alpha} \Pia\cdot\vb(\xa,t).
\label{defiEint}
\end{equation}
Due to the presence of dissipation and forcing $\FFa$ and $ \chi$, this interaction energy is not conserved in general. However, there may exist situations where dissipation and 
forcing balance statistically, so that the system reaches an out-of-equilibrium steady state.

\subsection{Weak \pin limit}
The equations of motions \eqref{interaction} takes a simple expression, in the "weak \pin" limit, where the intensity of the \pins are very small, $\gamma_\alpha\ll 1$ for any $\alpha$. In this case, 
$\Pia=8\nu\gamia/3$ and one can develop $\phiab^{-1}=(1+\gamib\cdot\rab/\|\rab\|)$. 
The equations of motions under such approximations are :
\begin{eqnarray}
\dot{\x}_\alpha&=&2\nu\sum_{\beta\neq \alpha} \left[ \frac{\gamib}{\|\rab\|}+\left(\gamib\cdot \rab\right)\frac{\rab}{\|\rab\|^3}\right],\nonumber\\
\dot{\pmb{\gamma}}_{\alpha} &=& 2\nu \sum_{\beta\ne\alpha}
\left[-(\gamia\cdot\gamib)\frac{\rab}{\|\rab\|^3} + 3\,(\gamia\cdot\rab)(\gamib\cdot\rab)\frac{\rab}{||\rab||^5} \right. 
\label{interactionweak}\\ && \left. - \, \gamia\frac{\gamib\cdot\rab}{\|\rab\|^3}+\gamib\frac{\gamia\cdot\rab}{\|\rab\|^3}\right]-\frac{6\pi\nu\psi(0) }{C_\psi\eta^2}\gamia+\frac{3E_\alpha}{8C_\psi\nu}\chi. \nonumber
\label{PWL}
\end{eqnarray}
These equations of motions are reminiscent of the equations of motions of the vortons (see Eq. \eqref{interactionvorton} in Appendix), with vectorial products being replaced by scalar product, and additional terms appearing. However, the motion and intensities of the \pins
are driven by forces decaying respectively like $1/r$ and $1/r^2$, rather than respectively $1/r^2$ and $1/r^3$ for the vortons. Moreover, the \pins are subject to a friction proportional to the
viscosity.\

The interaction energy in this case is:
\begin{equation}
E=\frac{16C_\psi\nu^2}{3\eta}\sum_{\beta\neq \alpha}\left[ \frac{\gamia\cdot\gamib}{\|\rab\|}+\frac{\left(\gamia\cdot \rab\right)\left(\gamib\cdot \rab\right)}{\|\rab\|^3}\right],
\label{Hinteraction}
\end{equation}
which is the classical self interaction energy of pair of singularities \cite{Pedrizzetti1992}.

\section{Dynamics of a pair of \pins}
\subsection{Interest of considering pair of \pins}
Previous experimental \cite{[S16],debue2021} and numerical investigations \cite{Nguyen20} about the location of the structures with extreme local energy transfer showed that they are located near interactions (possibly reconnection) of Burgers vortices. Previous and recent high-resolution numerical simulations of reconnection of anti-parallel vorticity filaments \cite{Kerr89,yao_hussain_2020} showed that the process is associated 
with the formation of a local cusp over each filament, that could possibly lead to a singular behaviour \cite{yao20}. This possibility was confirmed by a detailed study of the interaction of two Burgers vortices conducted by \cite{Kimura18, moffatt_kimura_2019,moffatt_kimura_2019_2,Moffatt20,Kimura_2017} using the Biot-Savart approximation. They showed that the interaction indeed leads to a cusp formation on the vortex line, with very large, possibly diverging velocities at the tip of the cusp, with orientation along the bisector of the angle of the cusp. The reconnection was also found to be associated with a depletion of the helicity,
that eventually reaches zero at the reconnection \cite{Kimura18}. \
In this picture, the associated \pins created by a coarse-graining at the Kolmogorov scale would then be located at the tip of each cusp, with spins in the direction of the bisector of the cusps (see Figure \ref{fig:geomdipole}-(a)). This shows the interest
of studying more closely the dynamics of a pair of \pins and see how it compares with known features of the reconnection. This is the aim of the present section. In all the sequel, we renormalize the length by $r_0$, the distance between the two \pins at time $t=0$ and the time by the associated viscous time $\tau_\nu = \r_0^2/\nu$. We first start with the simplest case, where the pair constitutes a dipole.

\subsection{Dynamics of a dipole of \pins}
\subsubsection{Equations}

\begin{figure}
\begin{center}
\usetikzlibrary{math} 

\tikzmath{\x = 2.8;\y = 2;\a=atan(\y/\x);\epsi = 0.5 ;\aepsi=\a+\epsi;\g= 1;\gm = \g/2.5; \n = 3;\xx=\x/\n;\yy = \y/\n;} 

\begin{tikzpicture}

	\draw[thick,densely dashed] (-\x,-\y)--(\x,\y);
	\begin{scope}[shift={(\x,\y)}]
	\draw[->, >=latex, thick] (0,0)--++(\xx,\yy) node[right] {$\pmb{e}_r$};
	\draw[->, >=latex, thick] (0,0)--++(\yy,-\xx) node[right] {$\pmb{e}_\theta$};
	\end{scope}

	\draw[->,>=latex, Red] (0,1)arc(90:\aepsi:1);
	\node at (0.4,1.2) {$\theta$}; 
	\draw[->, >=latex] (0,0)--(3,0) node[below] {$\pmb{x}$};
	\draw[->, >=latex] (0,0)--(0,3) node[left] {$\pmb{z}$};

\begin{scope}[shift={(\x,\y)}]
	\draw[->, >=latex,ultra thick] (0,0)--++(0,\g);
	\draw[fill=blue] (0,0)circle(0.05);
	\node[right] at (0.2,0) {$\pmb{x}_\alpha$};
	\node[left] at (0,\gm) {$\pmb{\gamma}_\alpha$};
\end{scope}
\begin{scope}[shift={(-\x,-\y)}]
	\draw[->, >=latex,ultra thick] (0,0)--++(0,-\g);
	\draw[fill=blue] (0,0)circle(0.05) node[left] {$\pmb{x}_\beta$};
	\node[right] at (0,-\gm) {$\pmb{\gamma}_\beta$};
\end{scope}		
\end{tikzpicture}
\put(-320,80){\includegraphics[width=0.49\textwidth]{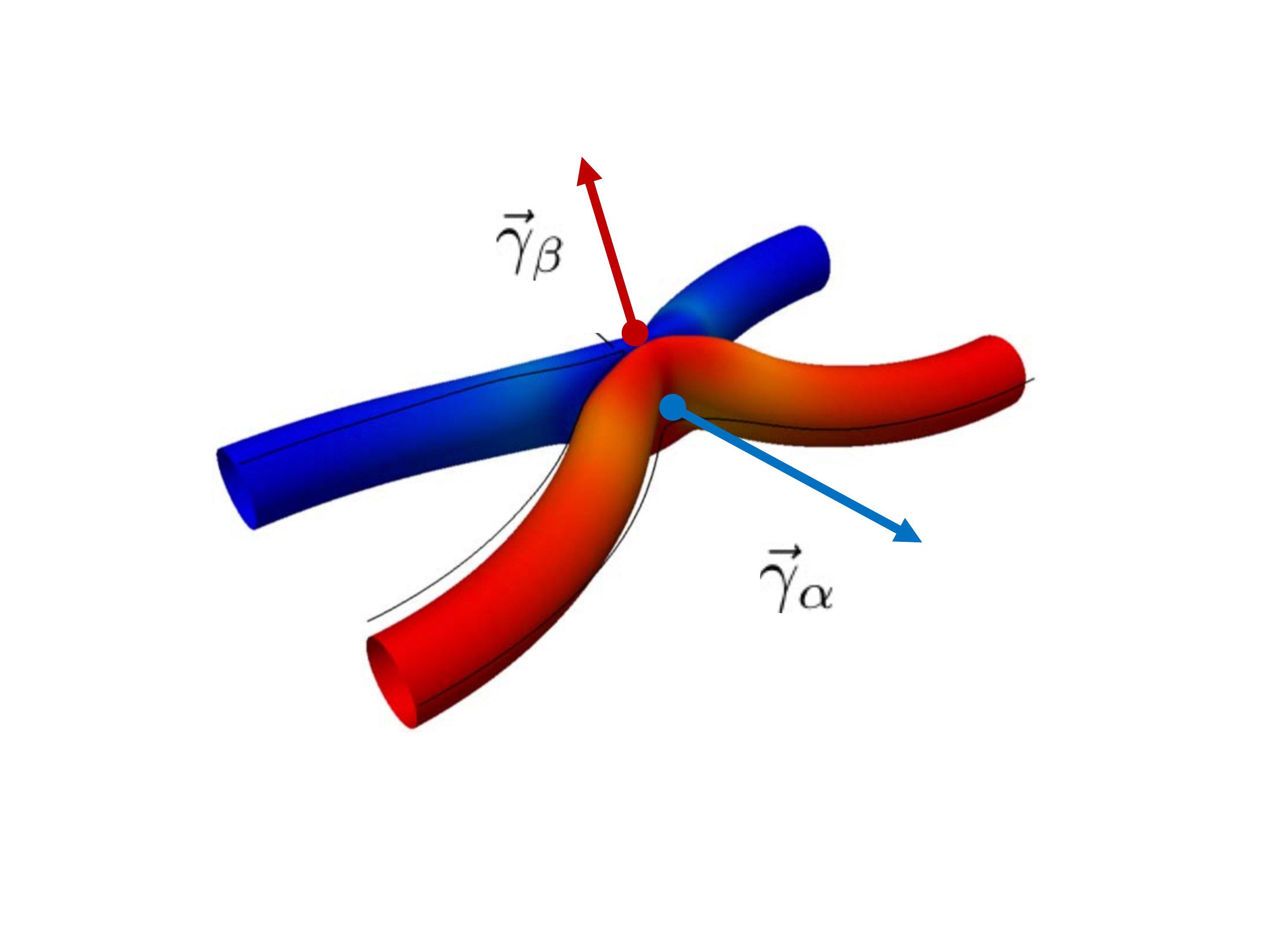}}
\put(-280,100){\bf(a)}
\put(-160,-20){\bf(b)}
\end{center}
\caption{(a) Schematic geometry of \pins creation at reconnection. (b) Geometry of the dipole: two \pins located at $\xa$ and $\xb$, and such that initially $\gamia+\gamib=0$. By convention, the angle $\theta$ is the angle between $\gamia$ and 
$\r=\xa-\xb$. The figure (a) is adapted from figure 3 of \cite{yao_hussain_2020}.}
\label{fig:geomdipole}
\end{figure}

Let us consider the dynamics of a dipole, sketched in Figure \ref{fig:geomdipole}-(b), made of two \pins located at $\xa$ and $\xb$, and such that initially $\gamia+\gamib=0$ and $\xa-\xb=r_0 \r$ . We have then $\vb(\xa)=-\va(\xb)\equiv \nu\V(\r)/r_0$ and $\FFa+\FFb=0$. Using the aforementioned non-dimensionalization, we then get the equation of motions:
\begin{eqnarray}
\dot{\x}_\alpha&=&\frac{\nu}{r_0}\V,\nonumber\\
\dot{\x}_\beta&=&-\frac{\nu}{r_0}\V,\nonumber\\
\dot{\pmb{\Pi}}_{\alpha}&=&\Pia\nabla_r \pmb{V}-\frac{\nu\psi(0)}{C_\psi}\left(\frac{r_0}{\eta}\right)^2\FFa+\frac{kT_\alpha r_0^2}{C_\psi\nu}\chi,
\label{interactionDipole}\\
\dot{\pmb{\Pi}}_{\beta}&=&-\Pib\nabla_r \pmb{V}-\frac{\nu\psi(0)}{C_\psi}\left(\frac{r_0}{\eta}\right)^2\FFb+\frac{kT_\beta r_0^2}{C_\psi\nu}\chi\nonumber.
\end{eqnarray}
Therefore, the center of mass of the dipole $\xa+\xb$ does not move, while the mean dipole strength $(\Pia+\Pib)/2$ obeys:
\begin{equation}
\dot{\pmb{\Pi}}_{\alpha}+\dot{\pmb{\Pi}}_{\beta}=\frac{(E_\alpha+E_\beta) r_0^2}{2C_\psi\nu}\chi.
\label{dipolestrength}
\end{equation}
We see that he forcing induces fluctuations proportional to the mean local energy $(E_\alpha+E_\beta)/2$ that destroy the dipole geometry over a viscous time scale. It therefore only make sense 
to study the dipole case in the low temperature limit where $(E_\alpha+E_\beta) /2\to 0$.

\subsubsection{Results at zero temperature}
Let us first investigate the dynamics in the zero temperature $E_\alpha+E_\beta=0$. In this case, the dipole remains exactly a dipole at all times and we have $\Pia=-\Pib\equiv \nu \pmb{\Pi}$, $\FFa=-\FFb\equiv~\FF$ and $\gamia=~-~\gamib\equiv~\gami$. The dipole dynamics of the quantities characterizing the dipole, namely $\r$ and $\gami$ (or equivalently $\pmb{\Pi}$),
can be obtained by taking the difference of the first two and the last two equations of Eq. \eqref{interactionDipole} to get:
\begin{eqnarray}
\dot{\r}&=&4\left(-\frac{\gami+\r/r}{r\phi_*}+(1-\gamma^2)\frac{\r}{r^2\phi_*^2}\right),\label{equa0} \\
\dot{\pmb{\Pi}}&=&\frac{2 \Pi}{r^3 \phi_*^3} A(\gamma,\theta) \, \pmb{r} - \frac{\psi(0)}{C_\psi}\left(\frac{r_0}{\eta}\right)^2\FF ,
\label{intdynamicsDipole}
\end{eqnarray}

where 
\begin{eqnarray}
A(\gamma,\theta) = \gamma \left(1 - 3\cos^2(\theta) - 3\gamma\cos(\theta) - \gamma\cos^3(\theta) - 2\gamma^2 \right)
\end{eqnarray}


where $r=\|\r\|$, $\cos(\theta)=(\gami\cdot \r)/(r\gamma)$, $\phi_*=1+\gamma\cos(\theta)$ and $\Pi=\|\pmb \Pi \|=G(\gamma)$. \

\begin{figure}
\begin{center}
\includegraphics[width=0.49\textwidth]{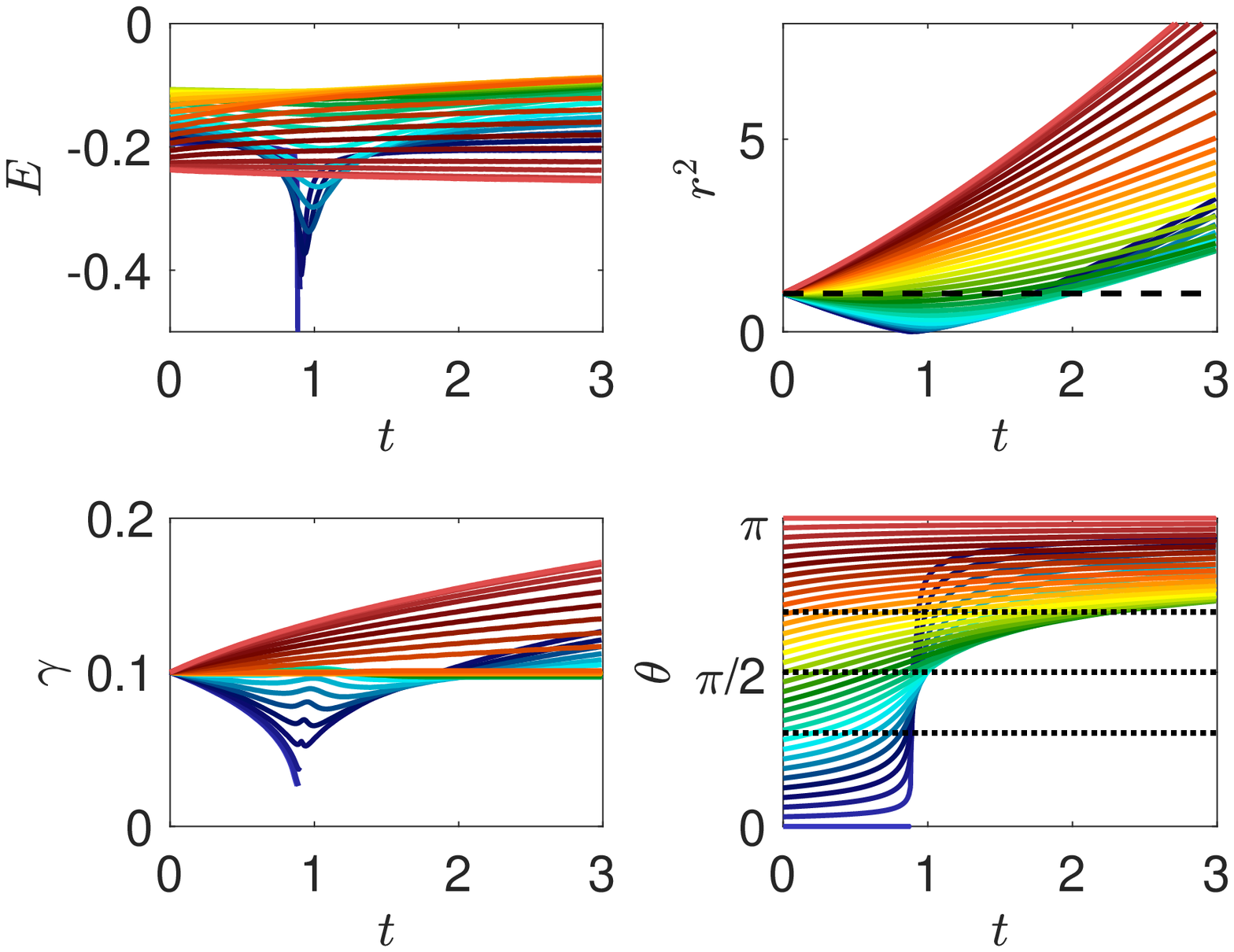}
\put(-95,75){\bf(a)}
\includegraphics[width=0.49\textwidth]{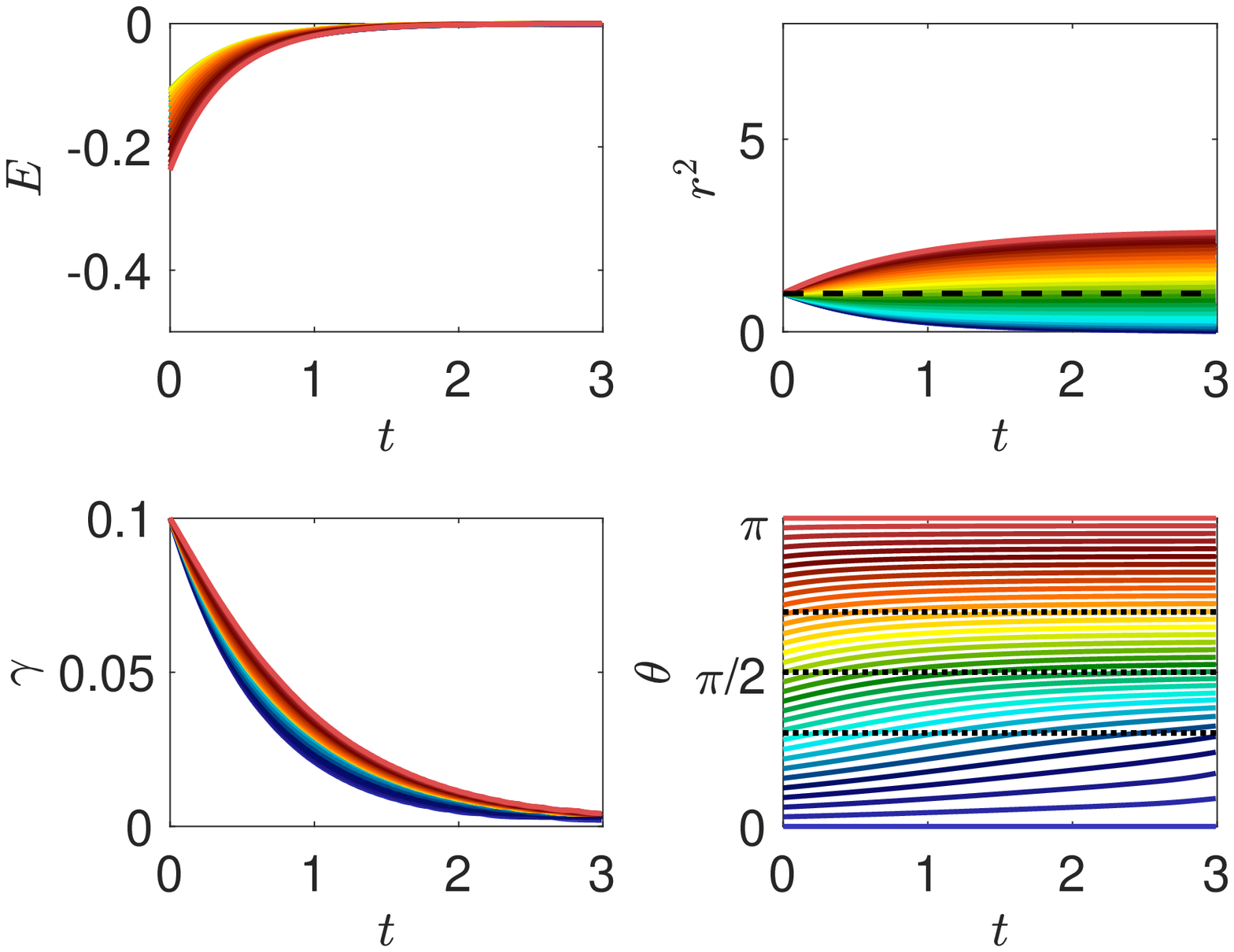}
\put(-95,75){\bf(b)}
\end{center}
\caption{Dynamics of a dipole of \pin for various initial conditions and (a) Without friction, corresponding to the initial stage of the dynamics, just after \pins creation; (b) With friction, corresponding to the late stage dynamics. The radius is initially fixed to $r=1$, the dipole intensity is initially set to $\gamma=0.1$ and the initial dipole orientation is fixed at different values between $0$ and $\pi$. The panel represent the time evolution of the different quantities: $E$: Interaction energy; $r^2$:~Square of Dipole separation; $\gamma$: Dipole intensity; $\theta$: Dipole orientation. Black dashed lines on $\theta(t)$ figure correspond to $\arccos \big(\frac{1}{\sqrt 3}\big),\frac{\pi}2$, and $\pi -\arccos \big(\frac{1}{\sqrt 3}\big)$.}
\label{fig:dynamique}
\end{figure}

\begin{figure}
\begin{center}
\includegraphics[width=0.49\textwidth]{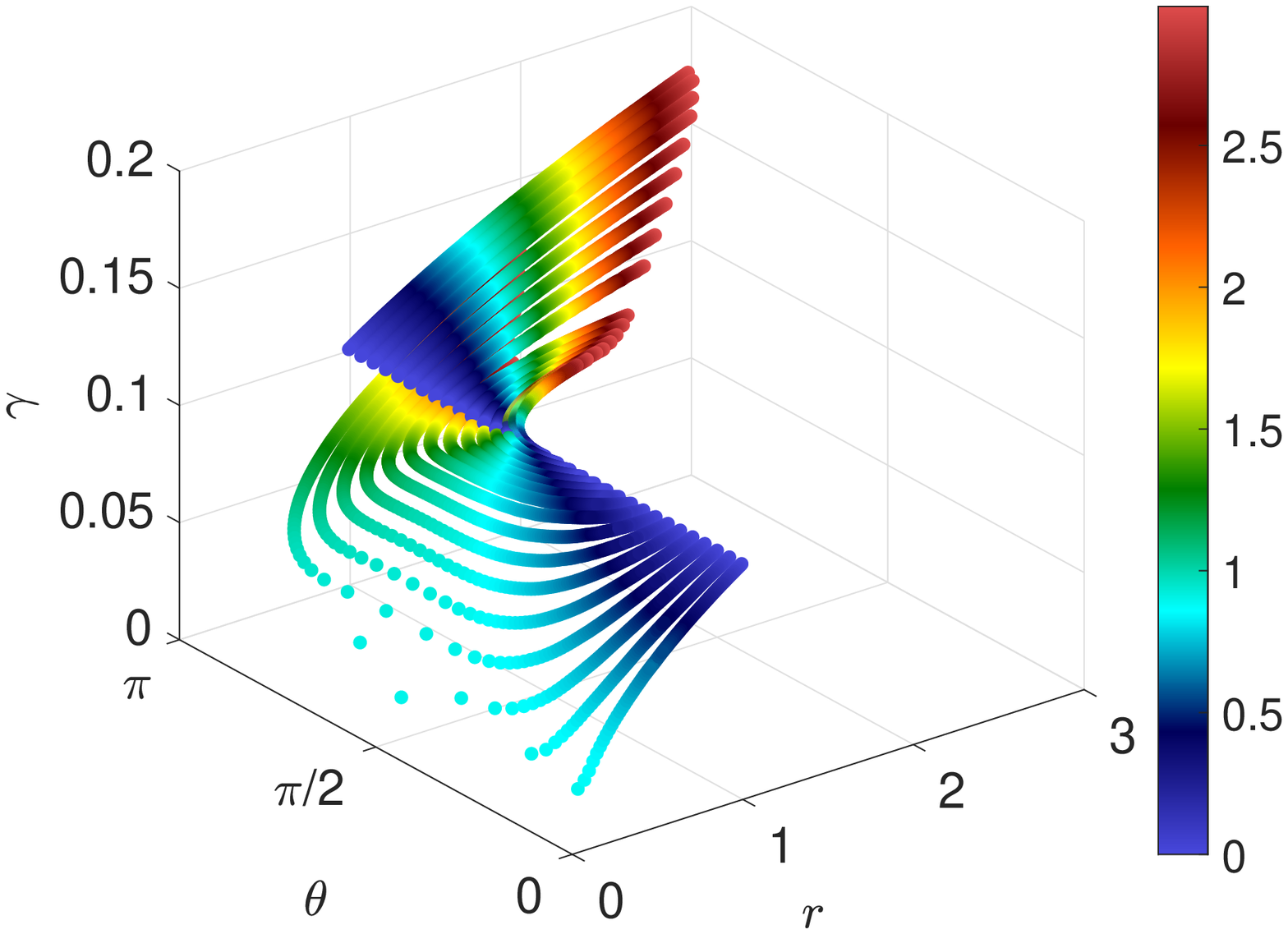}
\put(-160,120){\bf(a)}
\includegraphics[width=0.49\textwidth]{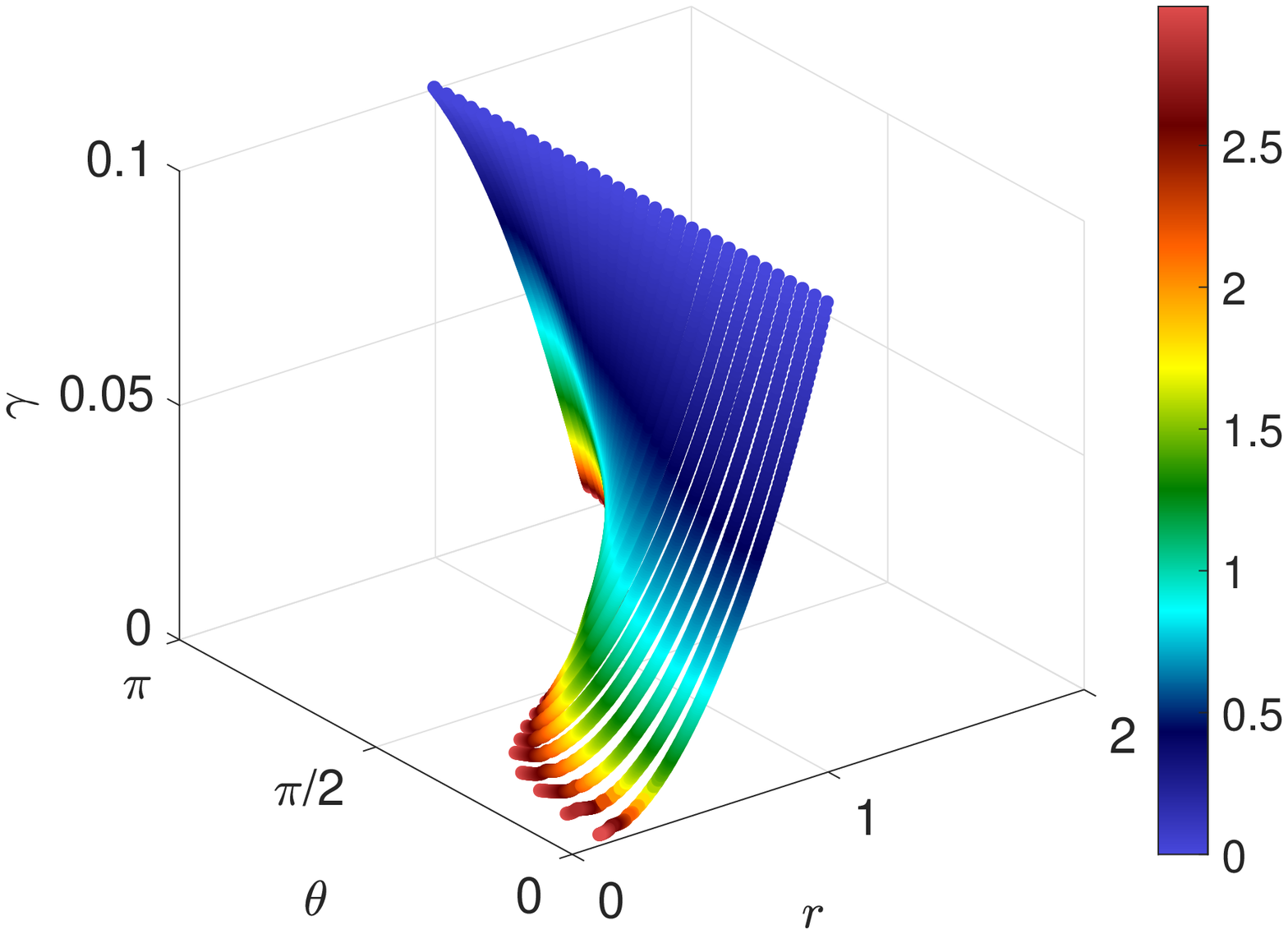}
\put(-160,120){\bf(b)}
\end{center}
\caption{Phase space for a dipole of \pins color-coded by the time, shown on the color bar. (a) Without friction, corresponding to the initial stage of the dynamics, just after \pins creation (b) With friction, corresponding to the late stage dynamics. The radius is initially fixed to $r=1$, the dipole intensity is initially set to $\gamma=0.1$ and the initial dipole orientation is fixed at different values between $0$ and $\pi$.}
\label{fig:attractor}
\end{figure}

From these expressions, we see that the evolution of $\r$ and $\pmb{\Pi}$ remains in the plane generated by the two vectors $\r$ and $\gami$. We are then left with only 3 independent quantities to determine the dipole axis and its orientation, namely $r$, $\theta$ and $\gamma$. The evolution of the first two quantities can be simply derived by projecting 
Eq. \eqref{equa0} on $\textbf{e}_r$ and $\textbf{e}_\theta$, while the last quantity can be obtained by taking the scalar product of Eq. \eqref{intdynamicsDipole} with $\pmb\Pi$ to get an evolution for 
 $\| \pmb\Pi\|^2$, which leads to $\gamma$ through $\gamma=G^{-1}(\| \pmb\Pi\|)$. We thus get after straightforward simplifications:
\begin{eqnarray}
\dot r&=& \frac{4}{r}\left(\frac{1-\gamma^2}{\phi_*^2}-1\right),\\
r\dot \theta&=&4\,\frac{\gamma \sin\theta}{r\phi_*}, \\
\dot{\pmb{\Pi}^2} &=&4\frac{\Pi^2}{r^2} A(\gamma,\theta)\cos(\theta)-\frac{2\psi(0)}{C_\psi}\left(\frac{r_0}{\eta}\right)^2\FF \cdot \pmb{\Pi}\label{holaop}
\end{eqnarray}
We have integrated the equations of motions \eqref{holaop} for fixed initial radius $r_0=1$ and $\gamma_0=0.1$ and various initial values of $\theta_0$ and taking $\psi(0)=1/(2\pi)^{3/2}$ 
(valid for $\psi$ Gaussian). The resulting evolutions are computed in two case, with and without friction. 
The first case corresponds to the initial stage of the dynamics, just after the \pins are created. Indeed, a \pin is created with an initial force corresponding to the local Reynolds stress $\FFa = \eta^3/(\nu^2 \psi(0))\nabla \cdot \tau^\eta(\xa)$, so that the dissipation is initially suppressed. In that case, we observe in Fig. \ref{fig:dynamique}-a that there are two
 fixed points of the dynamics for $\theta$: one stable and attractive, corresponding to $\theta=\pi$, and one unstable and repelling, corresponding to $\theta=0$. As a result, the \pins are mostly repelling each other, except when they start exactly anti-aligned and facing away each other, in which case they attract each other and anihiliate each other. The resulting dynamics can also be appreciated in the phase space, as shown in figure \ref{fig:attractor}-a.\
 
\begin{figure}

\includegraphics[width=0.5\textwidth]{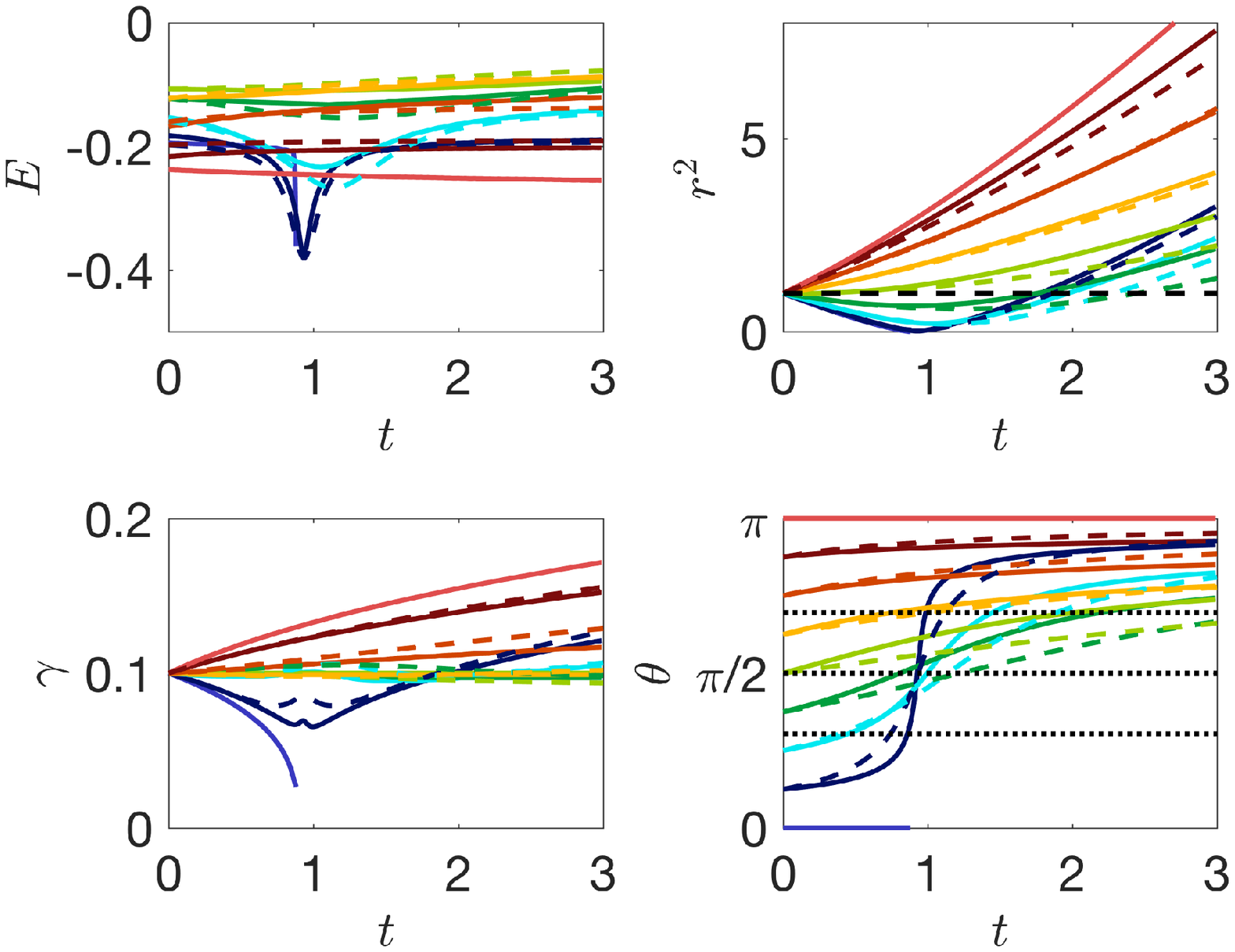}
\put(-95,75){\bf(a)}
\includegraphics[width=0.5\textwidth]{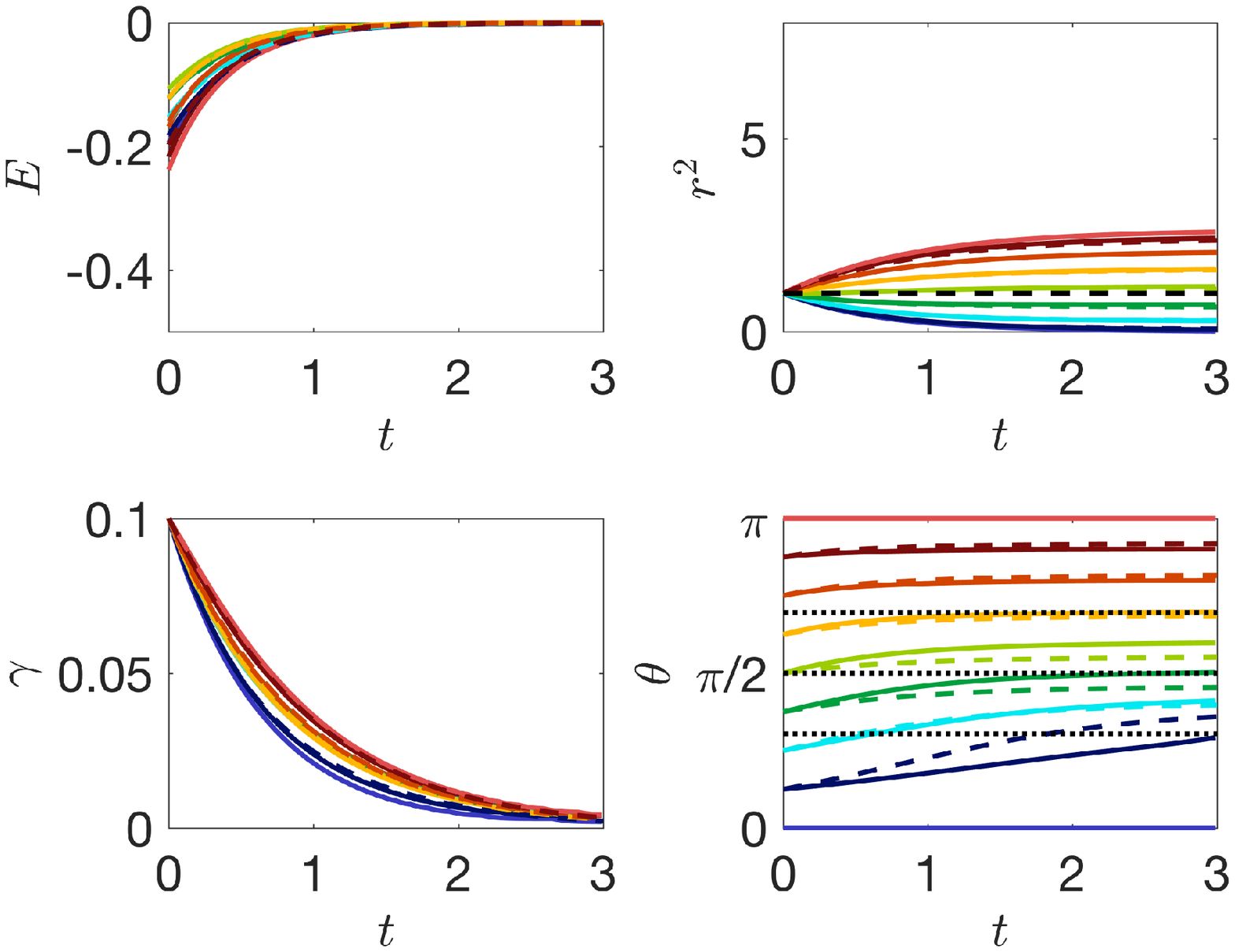}
\put(-95,75){\bf(b)}
\caption{ Comparison with the weak limit approximation: solid lines: complete model; dashed line:
weak limit; (a)Without friction. (b) With friction. The panel represent the time evolution of the different quantities: $E$: Interaction energy; $r^2$:~Square of Dipole separation; $\gamma$: Dipole intensity; $\theta$: Dipole orientation. Black dashed lines on $\theta(t)$ figure correspond to $\arccos \big(\frac{1}{\sqrt 3}\big),\frac{\pi}2$, and $\pi -\arccos \big(\frac{1}{\sqrt 3}\big)$.}
\label{fig:WLA}
\end{figure}

 Interesting scaling laws are observed in the two stages, that are reminiscent of what is observed during reconnection events
 The \pins with initial inclination in the interval
$[0,\pi/2]$ and different from $0$ start moving towards each other, while decreasing their strength and increasing their angle, in absolute value. Once they reach the value $\theta=\pi/2$, 
they change direction and get away from each other (see figure \ref{fig:WLA}-a). During the collapse stage, the radius of the dipole decreases approximately like like $\sqrt{t_c-t)}$, which is the Leray scaling \cite{Leray34}). The collapse stage is nearly universal, with weak dependance on the initial angle, while the escape depends more strongly on the initial orientation. Such asymmetry has been also observed in reconnection of quantum vortices \cite{Villois20}. During the separating stage, $\theta$ gets closer to $\pi$ and there is also an approximate power law escape law $r\sim \sqrt{t-t_c)}$. The scaling laws are explored further in the general case, in section 3.d.\

 At a later stage, the initial Reynolds stress has decayed and cannot balance anymore the \pins dissipation. The extreme case is when the Reynolds stress has decayed to zero, and when the dissipation is the strongest. This case has been studied by running another set of simulation including the friction, starting the dipole at $r_0=\eta$. It is shown
 in Figure \ref{fig:dynamique}-(b). We see that, as expected, the dissipation now induces a constant decline of the dipole intensity, that eventually results in the death of the dipole. 
 The resulting phase space is shown in figure \ref{fig:attractor}-b, where a plunging funnel corresponding to 
dissipation can now clearly be seen.
 There are now two situations depending on whether the initial angle is less or bigger than 
 $\pi/2$ . In the first case, the dipole contracts and orientates itself towards $\pi/2$ before its death. This situation can be associated to a reconnection event. For initial angles greater than $\pi/2$, the dipole expands
 while keeping its initial orientation before eventually dying and stopping.\

 Summarizing, the natural dynamics of a dipole without stochastic forcing is always dissipative, resulting in the final death of the dipole. Before its death, the dipole can experience either a first contraction stage of its initial angle is less than $\pi/2$, with dynamics resembling reconnection event, or experiences and expansion while it tries to anti-align its two components ($\theta\to \pi$).
 The question is now whether the dipole can be maintained for a longer time, and maintain another orientation if we take the stochastic forcing into account.
 As we showed in the beginning, as soon as we add some stochastic noise, the dipolar geometry breaks down in a viscous time scale. There is therefore a subtle interplay to be understood
 between the decay of the dipole, the forcing and the departure from dipolar geometry. There is however an interesting observation, that allows to tackle the problem in a simple 
 and elegant manner. Indeed, due to the friction, the \pin intensity decreases with time, and we are likely to enter into the weak \pin limit after a sufficient long time. The latter is much simpler to code and understand, since it
ressembles the dynamics of dipolar moments. In the sequel, we therefore investigate the finite but small forcing limit in the weak \pin approximation.

\subsection{Results at finite small temperature}
\subsubsection{From pair of \pins to dipole equations in the weak \pin limit}
 In such limit, we can obtain the dipole dynamics and the evolution of the dipole radius, orientation and strength from the evolution of three quantities: $R=r^2$, $C=\gami\cdot \r$ and
$N=\gami^2$. To obtain them, we first write the equations  Eq. \eqref{PWL} for two $\pins$ $\gamia$ and $\gamib$, and introduce three quantities: $\r=\xa-\xb$, ${\pmb \Gamma}=(\gamia+\gamib)/2$ and ${\pmb \Pi}=(\gamia-\gamib)/2$. By combination of the equations of motions, we obtain 3 coupled equations:
\begin{eqnarray}
{\dot \r}&=&-4\nu \left(\frac{{\pmb \Pi}}{r}+\r\frac{{\pmb \Pi}\cdot \r}{r^3}\right),\nonumber\\
{\dot {\pmb \Pi}}&=&2\nu\left[\left(\Pi^2-\Gamma^2\right)\frac{\r}{r^3}+3\frac{\r}{r^5}\left(({\pmb \Gamma}\cdot\r)^2-({\pmb \Pi}\cdot\r)^2\right)\right]-\frac{6\pi\nu\psi(0) }{C_\psi\eta^2}{\pmb \Pi}+\frac{3(E_\alpha-E_\beta)}{16 C_\psi\nu}\chi ,\\
\dot {\pmb \Gamma}&=&4\nu\left[{\pmb\Gamma} \frac{({\pmb \Pi}\cdot\r)}{r^3}-{\pmb \Pi}\frac{({\pmb \Gamma}\cdot\r)}{r^3}\right]-\frac{6\pi\nu\psi(0) }{C_\psi\eta^2}{\pmb \Gamma}+\frac{3(E_\alpha+E_\beta)}{16 C_\psi\nu}\chi\nonumber.
\label{usefuldp}
\end{eqnarray}
The last equation of Eq. \eqref{usefuldp} shows that ${\pmb \Gamma}$ is forced by $\bar E=(E_\alpha+E_\beta)/2$. If we start with a dipole condition $\Gamma=0$. 
In the small temperature limit $\bar E\ll 1$ and for time scale short with respect to the diffusive time scale $\tau_\nu=r_0^2/\nu$, we can then ignore
all the terms proportional to $\Gamma$ in the second equation of Eq. \eqref{usefuldp}. We then multiply the first equation by $\r$ and the second equation by ${\pmb \Pi}$ to obtain equations 
for $R$ and $N$, and we sum the first equation multiplied by ${\pmb \Pi}$ with the second equation multiplied by $\r$ to get the equation of evolution for $C$. After non-dimensionalization 
by $r_0$ and $r_0^2/\nu$ and rearrangement, we then obtain the 3 coupled equations:
\begin{eqnarray}
{\dot R}&=& -16 \frac{C}{R^{1/2}},\nonumber\\
{\dot C}&=& -2\left[ \frac{N}{R^{1/2}}+5 \frac{C^2}{R^{3/2}}\right]- \rho C+R^{1/2}\mu\zeta,\\
{\dot N}&=& 4\left[N \frac{C}{R^{3/2}}-3 \frac{C^3}{R^{5/2}}\right]- \rho N+N^{1/2}\mu\zeta,\nonumber
\label{jolieq}
\end{eqnarray}
where $\zeta$ is a delta correlated white noise obeying $<\zeta(t)\zeta(t')>=\delta(t-t')$, and the friction $\rho$ and forcing $\mu$ coefficient are given by 
\begin{equation}
\begin{array}{rcl}
\mu&=&\frac{3r_0^2(E_\alpha-E_\beta)}{16\nu^2 C_\psi},\\
\rho&=&6\pi\psi(0) \frac{r_0^2 }{\eta^2 C_\psi}.
\end{array}
\label{bobo}
\end{equation}
From $N$, $R$ and $C$, we then get $r=R^{1/2}$, $\gamma=N^{1/2}$ and $\theta=\mathrm{\arccos}(C/(r\gamma )$.

To check the validity of the weak limit approximation, we show in Figure \ref{fig:WLA} the comparison between the full dynamics computed from Eq. \eqref{intdynamicsDipole}
and its weak limit Eq. \eqref{jolieq} in the zero temperature limit $\mu=0$ and without ($\rho=0$) and with friction. We see that the two dynamics coincide very well for most cases,
and that the approximation is better on the late stage, when there is friction. Indeed, the friction forces decay of the dipole intensity.\

The weak dipole limit actually helps us identifying special angles for the dipole dynamics. Indeed, from the last equation of Eq. \eqref{jolieq}, we see that the first term of the r.h.s cancels 
whenever $\xi(\theta)=\cos(\theta)\big(1-3\cos^2(\theta)\big)=0$, corresponding to the three angles in the interval $[0,\pi]$, namely $\theta=\pi/2$, $\theta=\arccos(1/\sqrt(3))$ and $\theta=\pi-\arccos(1/\sqrt(3))$.
Those angles are identified by black dotted lines in figures \ref{fig:dynamique}, \ref{fig:WLA}, \ref{fig:SWEAK_attractors}. When the forcing and dissipation vanish or balance, they correspond to special directions where the dipole intensity can remain stationary. We see that indeed, the angle $\theta= \pi/2$ partitions the dynamics since the angles $\theta(t)$ increase and the radius decreases if and only if $\theta(t)$ is smaller than $\frac{\pi}{2}$.
The two other angles do not seem to play a specific role in the zero temperature limit. However, the situation is different in the other situation, at finite temperature, as we now show.

\subsubsection{Dynamics at finite temperature}
\begin{figure}
\includegraphics[width=0.5\textwidth]{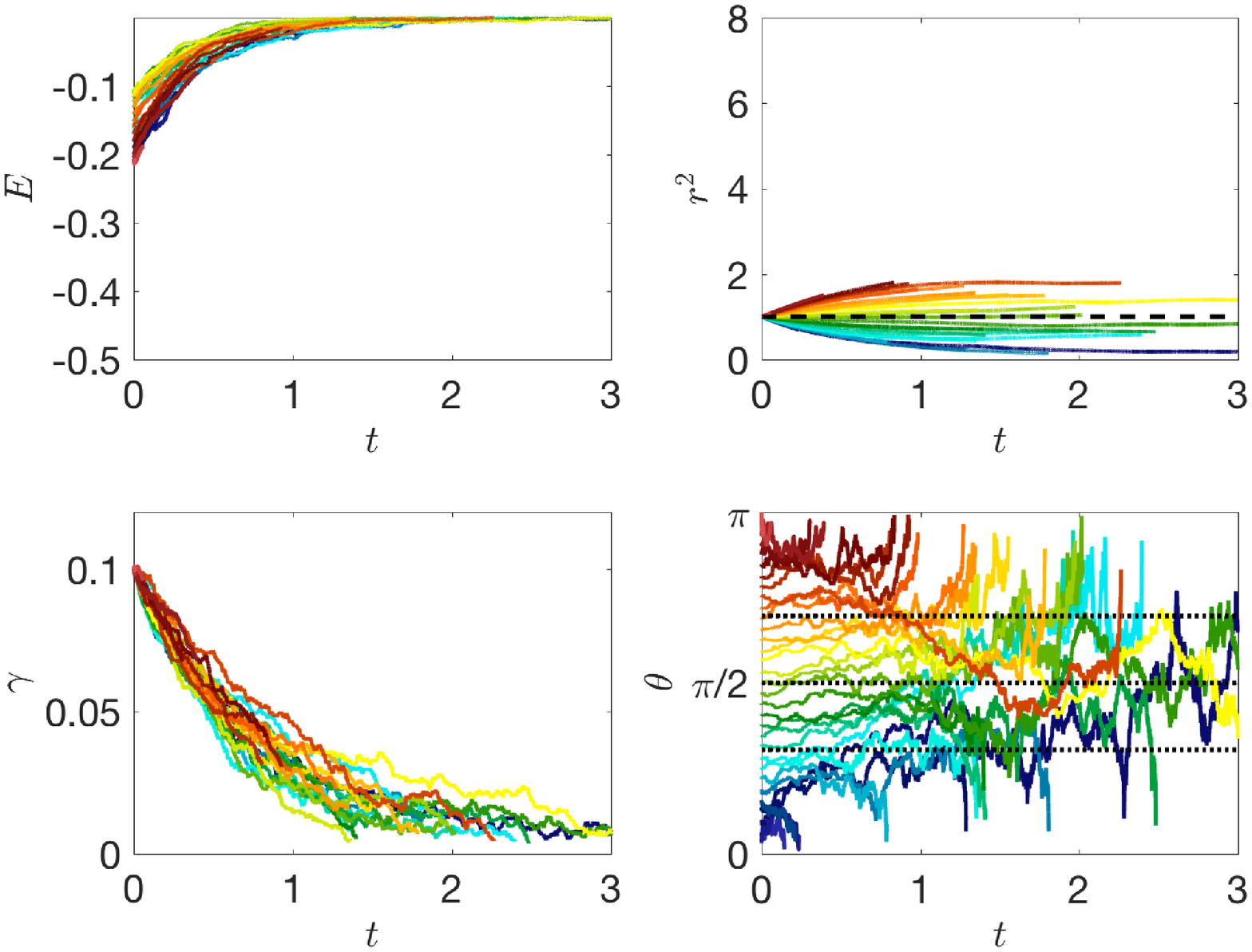}
\put(-100,130){\bf(a)}
\includegraphics[width=0.5\textwidth]{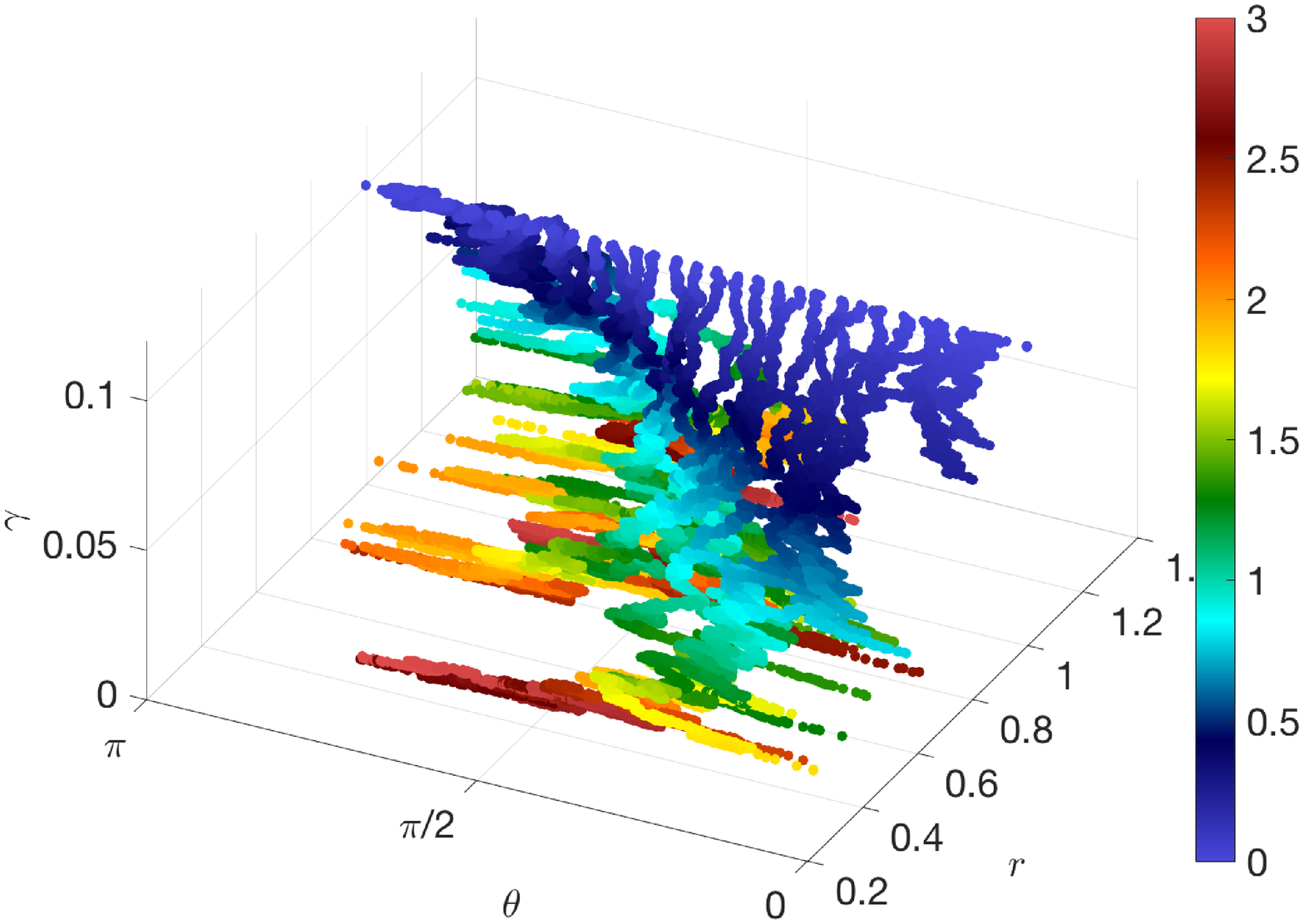}
\put(-60,130){\bf(b)}
\caption{Effect of noise in the weak limit dissipative case (a) on the dynamics. The panel represent the time evolution of the different quantities: $E$:Interaction energy; $r^2$ Square of Dipole separation; $\gamma$: Dipole intensity; $\theta$: Dipole orientation. Black dashed lines on $\theta(t)$ figure correspond to $\arccos \big(\frac{1}{\sqrt 3}\big),\frac{\pi}2$, and $\pi -\arccos \big(\frac{1}{\sqrt 3}\big)$; (b) on the phase-space, color-coded by the time, shown on the color bar.The radius is initially fixed to $r=1$, the dipole intensity is initially set to $\gamma=0.1$ and the initial dipole orientation is fixed at different values between $0$ and $\pi$. The intensity of the noise is $\mu=0.009$}
\label{fig:SWEAK_attractors}
\end{figure}

Integrating \eqref{jolieq} allows to efficiently study the dynamic of a noisy dipole, provided two criteria are satisfied:
i) $\gamma \leq 0.5$, in order to be in the weak limit and ii) $r$ is large enough ($r \geq \frac{1}{\sqrt \rho}$) so that the two \pins may still be considered as distinct
Moreover, for numerical reasons, we stop the integration whenever $\gamma\leq 0.01$ or $r \geq \sqrt \rho$, in which case we consider that the \pins are either dying or that the dipole has escaped to infinity. In practise, most of our simulations were stopped because either $\gamma\leq 0.01$ or $\gamma\geq 0.5$. 
Figure \ref{fig:SWEAK_attractors} shows the evolution a dipole satisfying of the equations of motions \eqref{jolieq} for fixed initial radius $r_0=1$ and $\gamma_0=0.1$, various initial values of $\theta_0$
and $\rho=0.12$ and $\mu=0.009$. 
Several tendencies emerge, as illustrated in figure \ref{fig:SWEAK_attractors}.
First the integration time is slightly longer as the noise may maintain $\gamma$ stationary for a certain amount of time. 
Second, the evolution of $\theta(t)$ is not monotonic anymore and the angle $\theta$ tends to get closer to the values cancelling $\xi(\theta)$, and tend to $\theta=\pi/2$ in most of the cases. As a result, the dipole can be maintain for some time
in a non-equilibrium balance, where dissipation exactly balances the stochastic forcing, allowing the dipole intensity to decay less rapidly, and the dipole to live longer.
In the phase space diagram ($r$, $\theta$, $\gamma$) trajectories with noise tends to allow the movement longer, keeping a larger $\gamma$ and leading to values of $\theta$ closer to $\pi$.
In the low noise regime, noise can therefore be a source of stimulation keeping the dipole alive and slowing down its death.
We have checked that when the noise is too large, then fluctuations are enough to bring the dipole intensity close to the limit $\gamma=0$ or $\gamma=1$ in a finite time, resulting in the dipole collapse or death quicker than in the zero temperature limit.

\subsection{General dynamics of a pair of \pins}
\subsubsection{Short time dynamics}
In the case of dipole, the short time dynamics corresponds either to an escape with $\theta \to \pi$ and $\gamma \to 1$ or first a contraction (close interaction) and than an escape. Here we consider the general case of a pair of \pins to determine whether such observation is robust or not. The relative dynamics is characterized in this case by 6 independent scalar variables which are the distance $r$ between the \pins, their intensities $\gamma_\alpha$ and $\gamma_\beta$, their angles $\theta_\alpha$ and $\theta_\beta$ defined by $\mathrm{cos}\, \theta_\alpha = (\gamia \cdot \r_{\alpha\beta})/(r\gamma_\alpha)$ and $\mathrm{cos}\, \theta_\beta = (\gamib \cdot \r_{\beta\alpha})/(r\gamma_\beta)$ and the angle $\varphi$ defined by $\mathrm{cos}\, \varphi = (\gamia \cdot \gamib)/(\gamma_\alpha \gamma_\beta)$. The dipole case studied in the previous section corresponds to $\gamma = \gamma_\alpha = \gamma_\beta$ , $\theta = \theta_\alpha = \theta_\beta$ and $\varphi = \pi$. We then integrate the dynamics with equations \eqref{interaction} without dissipation and noise for different initial conditions of these parameters, except that we always set $r_0 = 1$. Indeed since the \pin are created with an initial force corresponding to the local Reynolds stress $\FFa = \eta^3/(\nu^2 \psi(0)) \nabla \cdot \tau^\eta(\xa)$, the short time dynamics corresponds to the case without dissipation
and forcing.\par
Because we do not put any dissipation, we expect $\gamma$ to tend toward 1 as for the dipole, so we implement a stopping condition when there is one \pin for which $ \gamma > 1 -\varepsilon$, with epsilon a small parameter taken here to $\varepsilon = 10^{-2}$. We have ran the dipole dynamics with many different initial conditions, and identified 3 scenarios : 
\par
\begin{figure}
\includegraphics[width=0.49\textwidth]{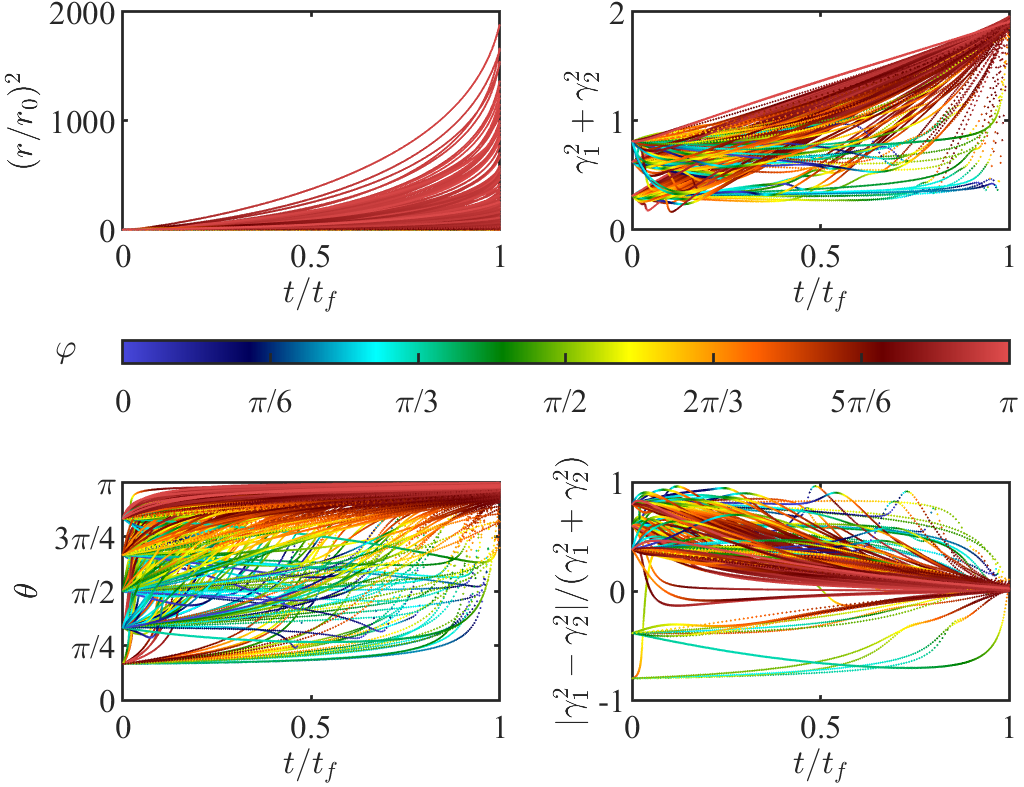}
\put(-91,88){\bf(a)}
\includegraphics[width=0.49\textwidth]{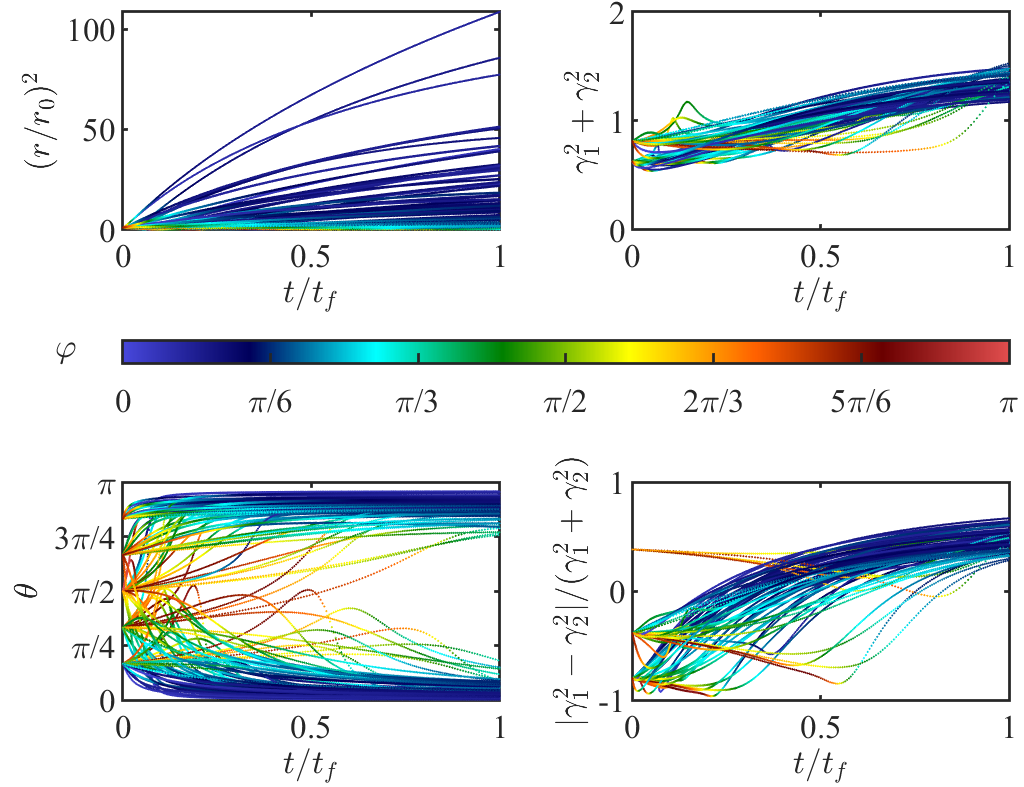}
\put(-91,88){\bf(b)}
\caption{Time evolution of the distance $r$, the \pins angles $\theta_1$, and $\theta_2$ (one line for each case), total intensity $\gamma_1^2+\gamma_2^2$ and 
anisotropy $(\gamma_1^2-\gamma_2^2)/(\gamma_1^2+\gamma_2^2)$ in the case expansion. (a) case of repelling dipolar expansion. (b) case of aligned expansion. The points are colored by the value of the pair mutual angle, $\varphi$.}
\label{fig:separation}
\end{figure}

 (i) repelling dipolar expansion, illustrated in figure \ref{fig:separation}-a. This case corresponds to the case where the two components run away from each other and gradually become
a repelling dipole: their mutual angle $\phi$ tends to $\pi$, while they become anti-parallel to their separation vector $\theta \to \pi$ and their intensities become equal to each other and tend to 1. In this case, the role of each \pins is symmetric. 

\par (ii) aligned expansion, illustrated in figure \ref{fig:separation}-b. In this case, one component grows larger than the other one, while both \pin become aligned with their separation vector and 
point in the same direction $\theta_1=\pi$, $\theta_2=0$, $\phi=0$. The component with the lower intensity moves faster and speeds ahead of the other one.

\par
\begin{figure}
\includegraphics[width=0.49\textwidth]{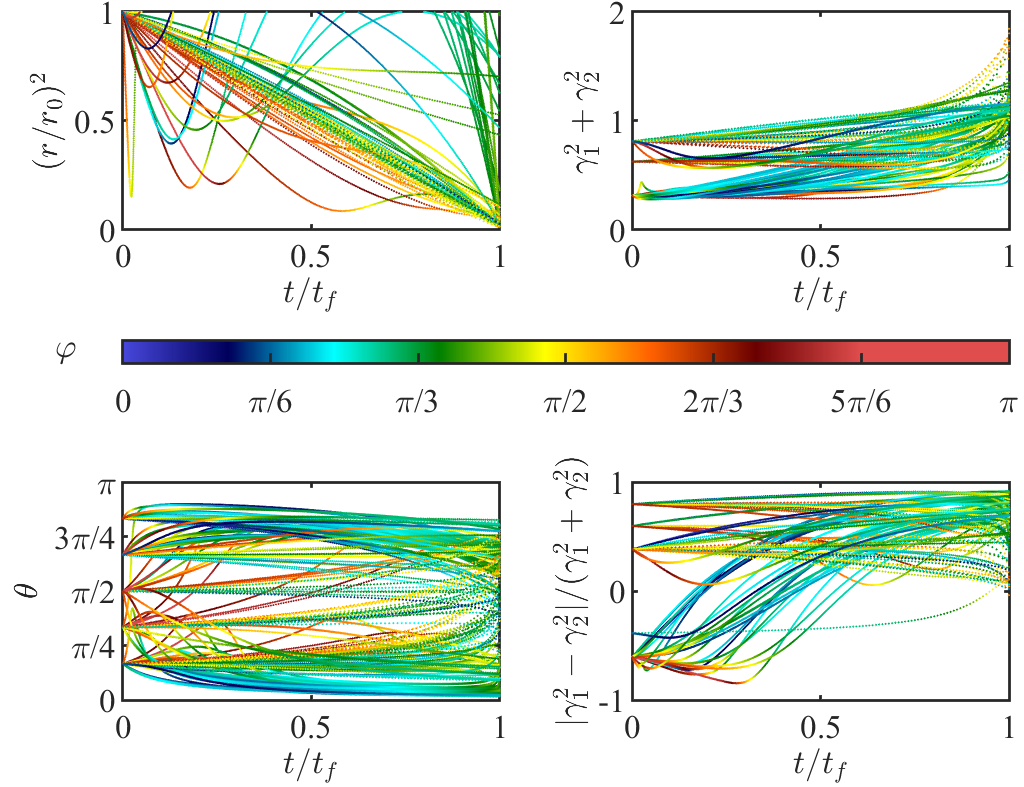}
\put(-91,88){\bf(a)}
\includegraphics[width=0.49\textwidth]{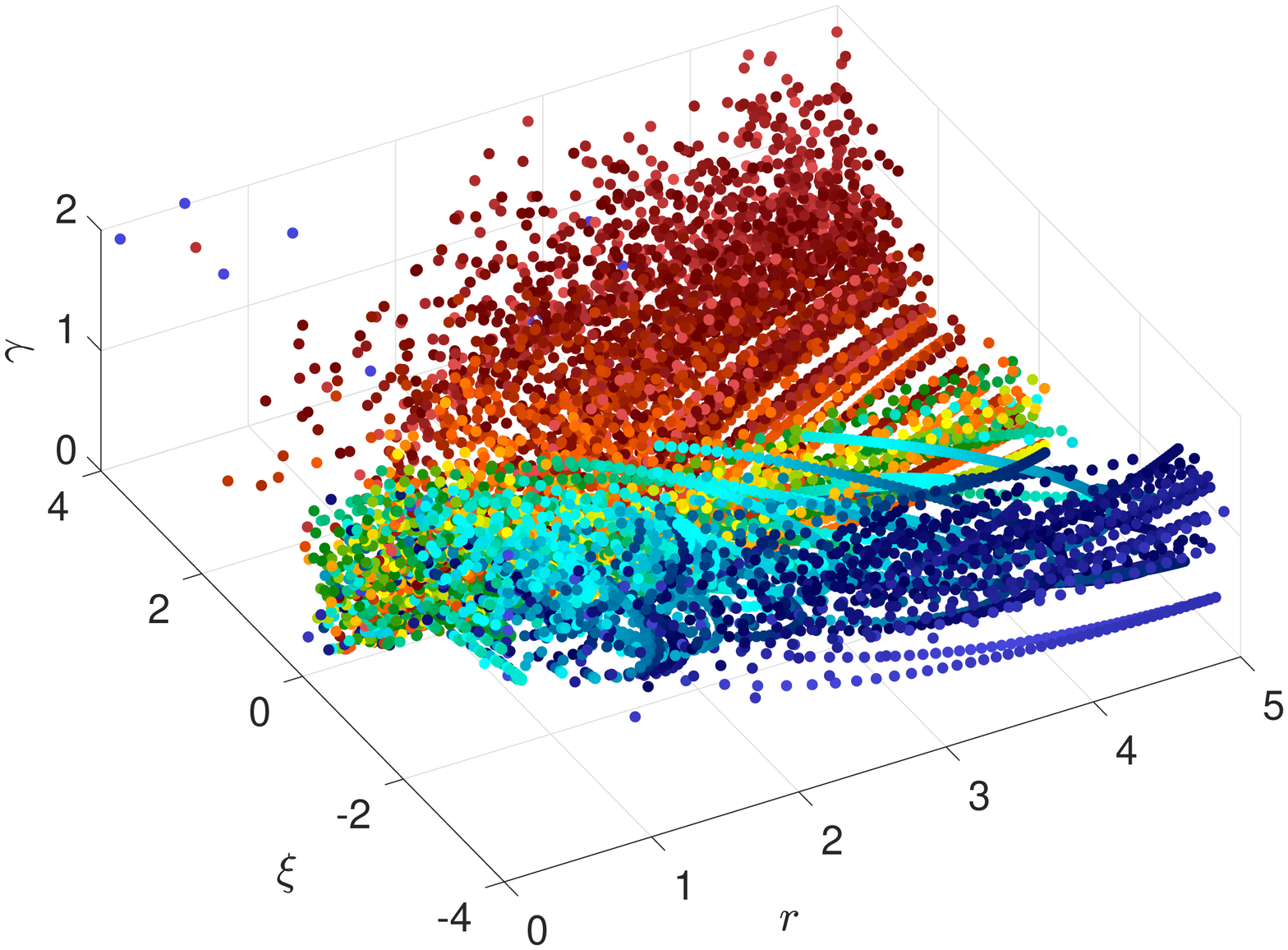}
\put(-91,130){\bf(b)}
\caption{(a) Time evolution of the distance $r$, the \pins angles $\theta_1$, and $\theta_2$ (one line for each case), total intensity $\gamma_1^2+\gamma_2^2$ and 
anisotropy $(\gamma_1^2-\gamma_2^2)/(\gamma_1^2+\gamma_2^2)$ in the case of explosive collapse. (b) 
Phase space for a pair of \pins in the initial stage of the dynamics, just after \pins creation. The phase space is $r$, $\gamma=\gamma_1^2+\gamma_2^2$ and $\xi=\xi(\theta_1)\xi(\theta_2)$.
The points are colored by the value of the pair mutual angle $\varphi$ in all the figures.}
\label{fig:collapse}
\end{figure}

 (iii) explosive collapse, illustrated in figure \ref{fig:collapse}-a. In this case, the two $\pins$ are attracted by each other, while one of the two $\pins$ rapidly reaches the asymptotic value
$\gamma=1$, corresponding to an infinite dissipation. In contrast with the expansion situations where the \pins tend to align or anti-align, the collapse case corresponds to intermediate values of $\varphi$ and $\theta$ different from $\pi$. This case can therefore be considered as the generic reconnection event.\

Interestingly, the different cases partition in different areas the phase space $r$, $\gamma=\gamma_1^2+\gamma_2^2$ and $\xi=\xi(\theta_1)\xi(\theta_2)$, with 
$\xi(\theta)=\cos(\theta)\big(1-3\cos^2(\theta)\big)$, as illustrated in figure \ref{fig:collapse}-b. One sees that the collapse mode tends to occur around $\xi=0$ meaning that at least one of the two \pins tend to orientate at $\pi/2$, $\arccos(1/\sqrt{3})$ or $\pi-\arccos(1/\sqrt{3})$ from the separation vector. In contrast, the two expansion modes proceed with $\xi=\pm 2$, corresponding to situations
where the \pins are aligned or anti-aligned with their separation vector.
Summarizing, we see that we observe two new cases with respect to the dipole dynamics, namely a new mode of expansion, made with two aligned dipole following each other, and 
a new mode of collapse following the law $\xi(\theta_1)\xi(\theta_2)=0$, with one component reaching the asymptotic value $\gamma=1$, and corresponding to a generic reconnection event. In the sequel, we study in more details these events.

\subsubsection{Scaling laws of collapse}
\begin{figure}
\includegraphics[scale=1.]{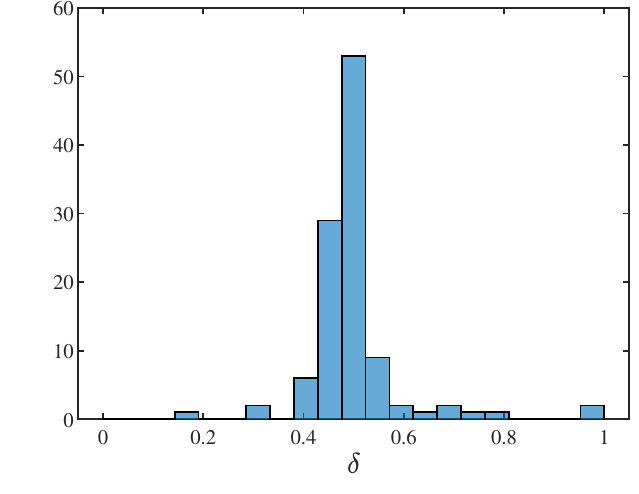}
\put(-20,130){\bf(a)}
\hfill
\includegraphics[scale=1.]{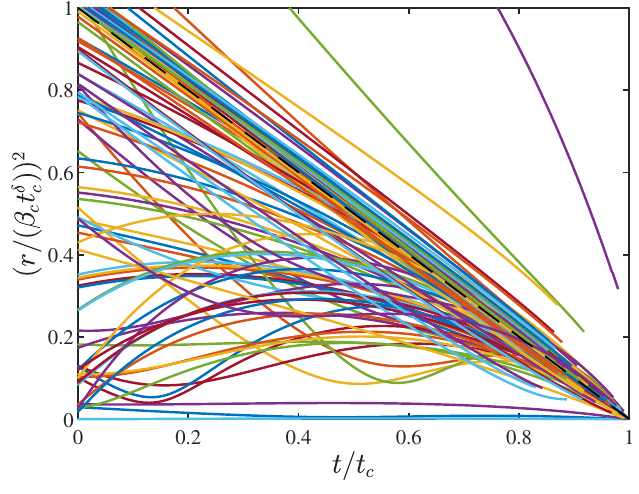}
\put(-20,130){\bf(b)}
\caption{(a) Histogram of the values of the power law exponent $\delta$. (b) Squared distance rescaled by for the explosive collapse cases. The black dashed line correspond to the Leray scaling with $\delta = 1/2$. We see that for $t/t_c$ close to $1$, most of the curves follow a power law with a power exponent $\delta$ close to $1/2$.}
\label{fig:result-fit}
\end{figure}

During the collapse stage, we fit the radius evolution with a power law in order to compare the exponent with the Leray scaling $\sqrt{t_c-t}$ \cite{Leray34}. Figure \ref{fig:result-fit} (a) shows the histogram of the values of the exponent obtained by fitting the law $r(t) = \beta_c (t_c - t)^\delta$ for the collapse cases. We see that most of the values are near $1/2$ which correspond to the Leray exponent. Figure \ref{fig:result-fit} (b) shows the time dynamics of rescaled squared radius. We see that the Leray scaling is verified asymptotically as $t/t_c \to 1$.

\subsubsection{Full collapse dynamics}
\begin{figure}
\centering
\includegraphics[width=1.\textwidth]{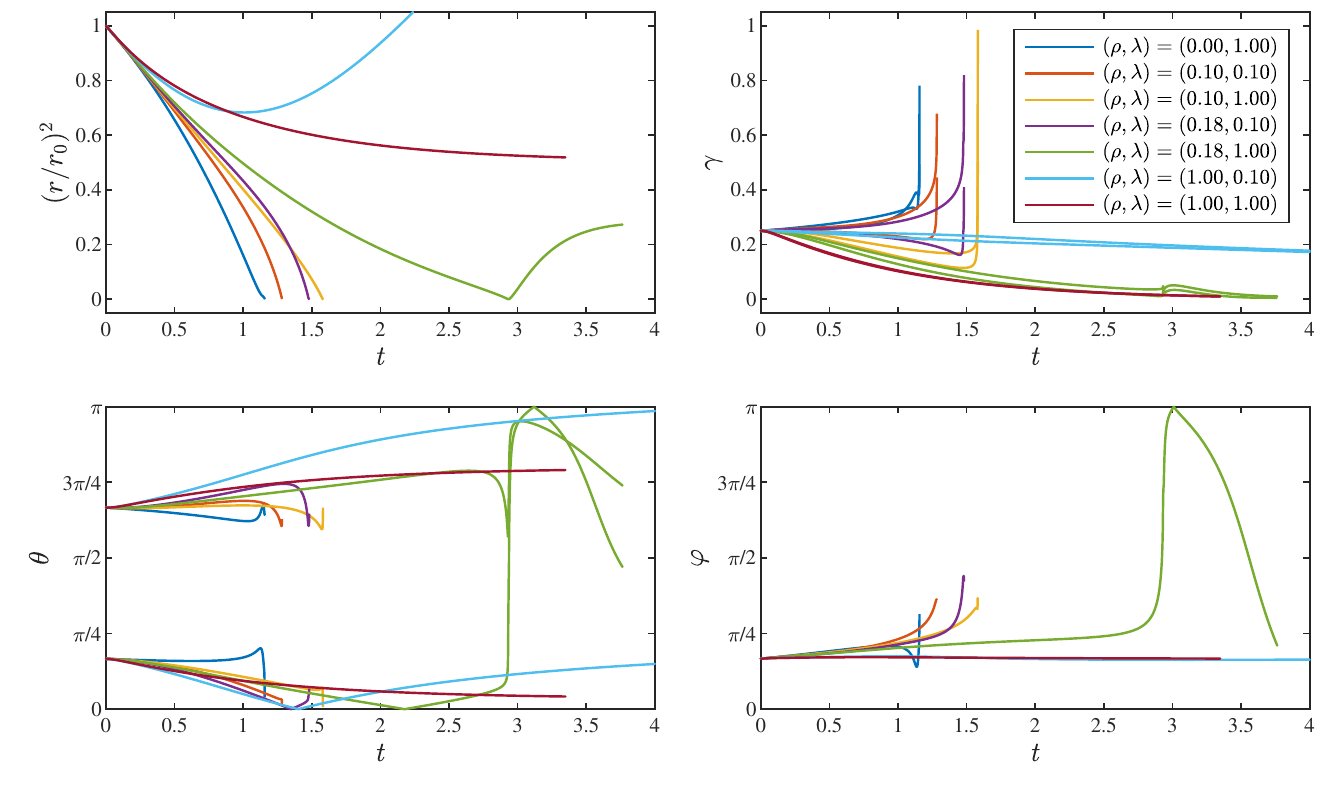}
\caption{Time evolution of the different variables characterizing the pair of \pins for 7 cases with the same initial configuration but with different forcing characteristic time coefficient $\lambda = \tau_\nu/\tau_\text{forcing} = r_0^2/(\nu \tau_\text{forcing})$ and dissipation coefficient $\rho = (\psi(0)/C_\psi)(r_0/\eta)^2$. We see that on the one hand, when dissipation is high and the forcing time is short, the \pins die very fast with no close interaction. On the other hand, if the dissipation is low, the dynamics is very similar to the case without dissipation, and the \pins still collapse in a explosive manner with their intensities tending to 1. Two intermediate cases are found where we have both the collapse dynamics and a separation dynamics without explosion.}
\label{patching}
\end{figure}

\begin{figure}
\includegraphics[scale=1.]{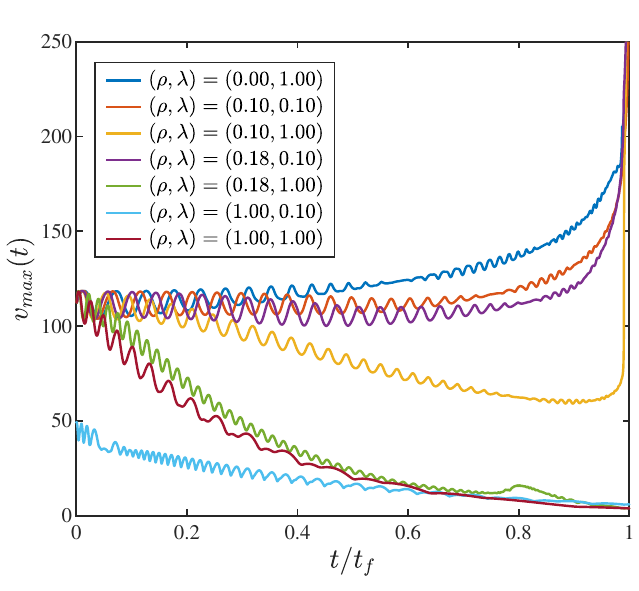}
\put(-20,145){\bf(a)}
\hfill
\includegraphics[scale=1.]{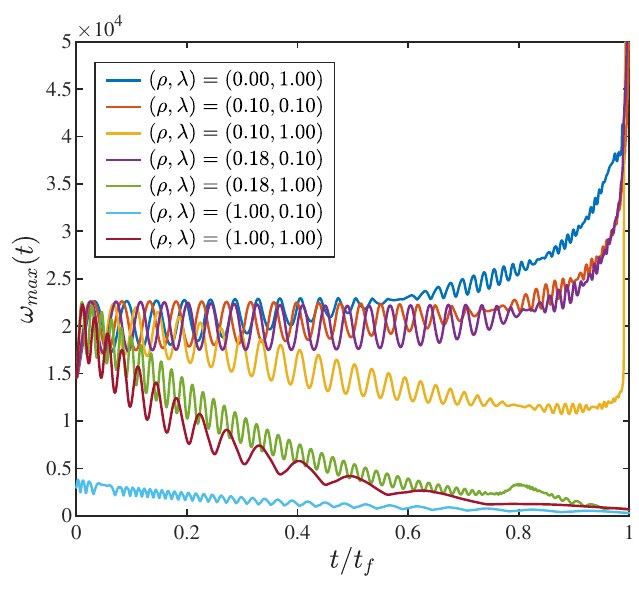}
\put(-20,145){\bf(b)}
\caption{Time evolution of the maximum of velocity field (a) and vorticity field (b) around a pair of \pins with same initial configurations and different forcing characteristic time coefficient $\lambda =\tau_\nu/\tau_\text{forcing}$ and dissipation coefficient $\rho$. The values correspond to a moving average over 15 time steps.}
\label{fig:ovmax}
\end{figure}

\begin{figure}
\centering
\includegraphics[scale=1.]{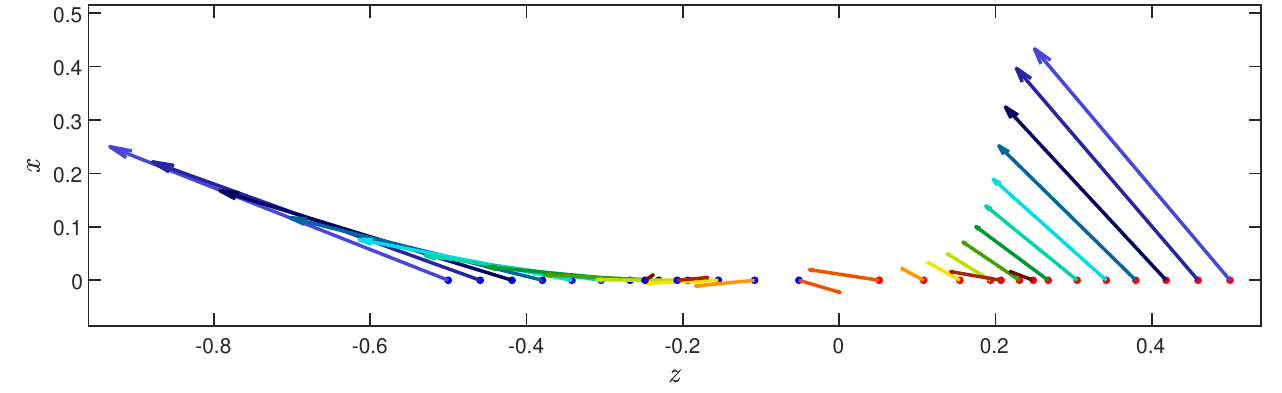}
\put(-25,100){\bf(a)}\\
\centering
\includegraphics[scale=1.]{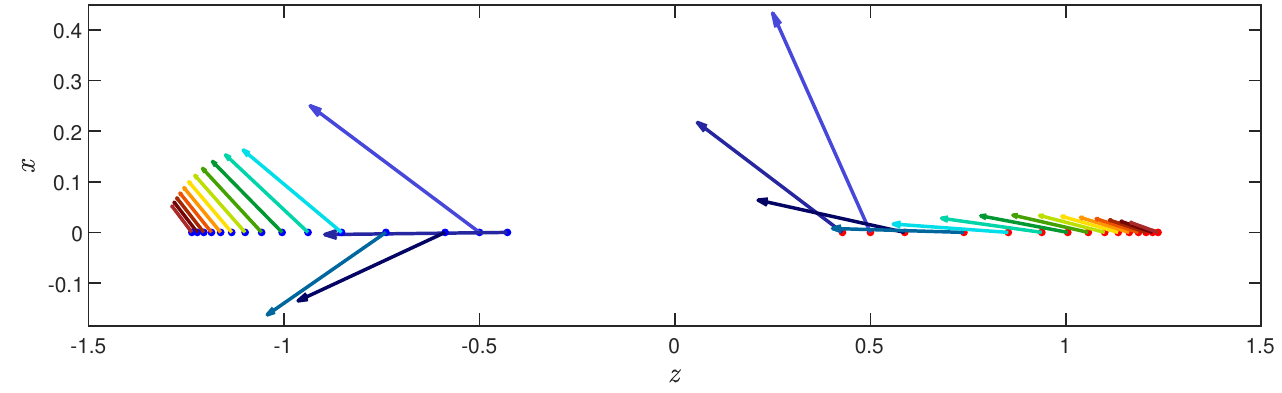}
\put(-25,100){\bf(b)}
\caption{Dynamics of a pair of \pins as a function of time for two different forcing characteristic time coefficient ($\lambda $) and dissipation coefficient $\rho$ : (a) $(\rho,\lambda) = (0.18,1.00)$ and (b) $(\rho,\lambda) = (1.00,0.10)$. The color of the vector codes the time, from $t=0$ (dark blue) to $t=t_{\text{final}}$ (dark red). In both cases, the two points of the pair (blue and red points) move initially towards each other and then their orientations change and they repulse each other. In the case of (a), the transition between the two phases has a configuration similar to the dipole with one of the \pins axis abruptly turning from an angle close to 0 to an angle close to $\pi$, then the \pins die very fast. In the case of (b), the dynamics is smoother with hardly any change in the dipole relative orientation, only the axis angles change slowly and the \pins survive a long time with a stable configuration.}
\label{fig:dip-dyn}
\end{figure}

The full dynamics of reconnection events can be investigated using a patching between the short time behavior, and the large time behavior, allowing e.g. the turbulent stress to decay like $\exp(-t/\tau_\text{forcing})$, where $\tau_\text{forcing}$ is a time scale associated to large scales. As shown on Figure \ref{patching}, for the same initial configuration, when dissipation is high and the forcing characteristic time is short, the \pins die very fast with no close interaction. On the contrary, if the dissipation is too low with a forcing persistent enough, the dynamics is very similar to the case without dissipation, and the \pins still collapse in a explosive manner with their intensities tending to 1. Two intermediate cases are found where we have a collapse stage followed by a separation stage without explosion. These typical examples are illustrated in figure \ref{fig:dip-dyn} (a-b). In both cases, we observe first a collapse phase and then a separation phase although the particular dynamics are quite different. In the case of (a) where the dissipation and the forcing characteristic time are rather small, the transition between the two phases happens at a closer distance and has a configuration similar to the dipole with one of the \pins axis abruptly turning from an angle close to 0 to an angle close to $\pi$, then the \pins die very fast. In the case of (b) with larger dissipation and a more persistent forcing, the dynamics is smoother with hardly any change in the dipole relative orientation, only the axis angles change slowly and the \pins survive a long time with a stable configuration. During the interaction, the maximum velocity and vorticity near the pair of \pins shown on Figure \ref{fig:ovmax} exhibit marked oscillations, due to the finite resolution of the grid. Using a moving average, we see however that for the cases where the intensities tend to 1, the maximum velocity and vorticity tend to infinity as expected. If we now look at the case of close interaction with a final separation corresponding to Figure \ref{fig:dip-dyn} (a), they first decrease during the collapse stage until the time of minimum of $r$, after which they increase until the angle is close to $\phi= \pi$, then, they finally both decay to zero when separating. This behaviour is reminiscent of what is happening during a reconnection of vortex rings, where the distance between rings decay like $\sqrt{t_c-t}$, with maximum velocity and vorticity growing up then decaying \cite{yao2020}. \
 
\section{Discussion}
We have introduced a model of singularities of Navier-Stokes, named \pins, that are discrete particles characterized by their position and "spin". These particles follow a nontrivial dynamics, obtained by the condition that the coarse-grained velocity field around these singularities obeys locally Navier-Stokes equations. We have shown that this condition can only be satisfied provided the coarse-graining scale is of the order of the Kolmogorov scale.When immersed in a regular field, the \pins are further 
transported and sheared by the regular field, experiencing a friction together with an energy injection coming from by Reynolds stress of the regular field. We have used these properties
to {\sl devise a toy model} of \pin interactions, that can be either seen as a generalization of the vorton model of Novikov\cite{Novikov83}, that was derived for the Euler equations, or as 
an Ising model of Navier-Stokes singularities, that can be use to get insight on small scale behavior of Navier-Stokes equations. As an example, we have studied the properties of the interaction of two \pins, at the early and late stage of their evolution, and in the presence or absence of a stochastic forcing induced by the possible Reynolds stress.\

Quite interestingly, we have identified several modes of interactions at short times, that are characterized by the values of the parameter $\xi(\theta)=\cos(\theta)\big(1-3\cos^2(\theta)\big)$, where 
$\theta$ is the angle between the spin of the \pin and the axis of the pair. Specifically, in the absence of noise, we identified two modes of expansion of the pair with $\xi=\pm 2$, corresponding to situations where 
the \pin are aligned or anti-aligned with their separation vectors, and one mode of collapse with $\xi=0$. In the presence of noise, we observe and additional transient non-equilibrium steady state expansion mode, 
with $\xi=0$, and the \pins are perpendicular to the axis of the pair. The quantity $1-3\cos^2(\theta)$ actually plays an important role in the theory of liquid crystals, as its average defines the order parameter of the system $s=(1-3\cos^2(\theta))/2$, with possible transitions between liquid ($s=0$) and nematic phase ($s=1$). The different interaction modes therefore open the 
way to interesting different collective behaviors when considering a larger collection of \pins. Whether such behaviors are of relevance to the actual physics of turbulence is still an open issue, as
the \pin model ignores on a number of issues that may limit its range of validity: existence of large nonlocal energy transfer at the Kolmogorov scale, dilute approximation for the \pin, scale separation between the \pin and the ambiant large scale velocity field, to name but a few.\

Our study of the interaction of two \pin however already revealed some interesting similarities with reconnection between two vortex rings. Indeed, we observe that the 
 collapse generally obeys a $\sqrt{t_c-t}$ that is observed during reconnection, and is characterized by transient growth of velocity or vorticity like in the reconnection. From another point of view, 
the \pins dynamics is also reminiscent of the two fluid model of superfluid, where the "regular" field, made of phonons, interact with the local topological defects that form the quantized vortices.
However, as shown by \cite{Villois20}, the interaction of quantized vortices leads to Leray scaling, with distance between vortices decaying like $\sqrt{t_c-t}$.
In a broader sense, the description of the interaction between \pins and a regular field is parallel to the interaction of localized wave packets interacting with a mean flow, in the WKB-RDT model of 
\cite{Laval2004}. By analogy, one may then wonder whether it would be possible to use the \pins as a subgrid scale model of turbulence, allowing to describe the interaction of a velocity field filtered at the Kolmogorov length, with a collection of \pins that encode the very intense energy transfers that are observed when scanning very small scales of turbulence \cite{[S16]}. If the \pin model proved accurate enough to describe small scale turbulence, it would then enable the use of larger time-steps, as the motion of the small scale motions is governed by Lagrangian motions. Another issue is whether a short range 
regularization is needed at short distances to make the model applicable to subgrid modelling.\

Even if the \pin model does not accurately describe the behavior of small scale turbulence, it is is an out-of-equilibrium statistical model of Navier-Stokes singularities with many interaction modes, that bears some similarity with liquid crystal interactions. It may then stimulate new ideas regarding turbulence dynamics and properties and play a similar role than the Ising model in statistical mechanics.

\enlargethispage{20pt}





{This work has been funded by the ANR, project EXPLOIT, grant agreement no. ANR-16-CE06-0006-01.}

{We thank I. Dornic for useful discussions and J. Gibbon and Y. Pomeau for useful comments. The constructive remarks of two referees led to substantial enrichment of the model with respect to its original setting.}


\vskip2pc

\bibliographystyle{RS}
\bibliography{pincon} 

\newpage
\section{Appendix}

\subsection{Useful properties}
We introduce the function:
$\phi(\x,\gamma)=\|\x\|-\gamma\cdot\x$. Such function has the properties:
\begin{eqnarray} 
\grad_{\x}\phi&=&\frac{\x}{\|\x\|}-\gamma\\
\phi&=&\x\cdot\grad_\x\phi\\
\grad_{\gamma}\phi &=&-\x\\
\Delta_\x(\ln(\phi))&=&\frac{1-\gamma^2}{\phi^2}.
\end{eqnarray}
Therefore, $\va$ can also be written:
\begin{equation}
\va=-2\grad(\ln\phi_\alpha)+2\x \Delta \ln(\phi_\alpha).
\label{propSympa}
\end{equation}
With such expression, it is easy to check that $\va$ is of zero divergence everywhere except at $x=0$, where it is undefined.

\subsection{Computation of the generalized momentum}
By definition:
\begin{equation}
\Pia=\frac{1}{4\pi}\oint_{S_{\xa}}\va {dS},
\label{defimom}
\end{equation}
where ${S_{\xa}}$ is a sphere of center $\xa$, and of radius unity, and the integration is perform only over the surface of the sphere. Taking spherical coordinate system with respect to a vertical axis along $\gamia$, $\x-\xa=(\cos\psi\sin\theta,\sin\psi\sin\theta,\cos\theta)$, it is easy to see that the azimutal average of $\x-\xa$ perpendicular to $\gamia$ is zero, and that the only nonzero component is along $\gamia$, and gives $\langle\x-~\xa\rangle_\psi=(0,0,\cos\theta)$. Using the fact that $\cos\theta=(1-\phi_\alpha)/\gamma$ with $\gamma=\|\gamia\|$, we thus get the azimutal average of $\va$ as
\beaa
\langle\va\rangle_\psi&=&(0,0,C);\nonumber\\
C&=&\frac{2\gamma}{\phi}-\frac{2(1-\phi)}{\gamma\phi}+2\frac{(1-\gamma^2)(1-\phi)} {\gamma \phi^2},
\eeaa
where we have dropped the subscripts $\alpha$ for simplicity.
We may easily compute the integration of the various term over $\theta$ since after a change of variable $y=\cos\theta$, and we get for any $n\ge 0$
\beaa
\langle\frac{1}{\phi^{n+1}}\rangle_\theta&=&\int \frac{\sin\theta}{\phi^{n+1}}d\theta;\nonumber\\
&=&\int_{-1}^{1}\frac{dy}{(1-\gamma x)^{n+1}}\nonumber\\
&=&\frac{1}{n\gamma}\left(\frac{1}{(1-\gamma)^n}-\frac{1}{(1+\gamma)^n}\right),
\eeaa
with the convention that $1/n x^n=\ln(x)$ when $n=0$. Summing all the terms, we finally obtain equation \eqref{GenMom}.

\subsection{Velocity gradient tensor}
\beaa
\nabla \textbf{U} &=& \left( \frac{\partial U_i}{\partial x_j} \right)_{i,j}\\
&=& -\frac{2}{\phi}\left[ \left( \pmb{\gamma} - \frac{\textbf{x}}{\lVert \textbf{x}\rVert} + 2 \frac{(1-\gamma^2)}{\phi}\textbf{x}\right) \otimes \frac{\nabla \phi}{\phi} + \left( \frac{1}{\lVert \textbf{x}\rVert} - \frac{1-\gamma^2 }{\phi} \right)\mathbb{I} - \frac{\textbf{x} \otimes \textbf{x}}{\lVert \pmb{x}\rVert^3} \right] \\
&=& -\frac{2}{\phi^2}\left[ -\pmb{\gamma} \otimes \pmb{\gamma} + \frac{\left(3\gamma \mathrm{cos}(\theta) -\gamma^2(2+\mathrm{cos}^2 (\theta)) \right)}{(1 - \gamma \, \mathrm{cos}(\theta))}\frac{\textbf{x} \otimes \textbf{x} }{\lVert \textbf{x}\rVert^2} + \frac{\pmb{\gamma} \otimes \textbf{x}}{\lVert \textbf{x} \rVert} \right. \\
&& \left. -\frac{\left(\gamma \, \mathrm{cos}(\theta) + 1-2\gamma^2\right)}{(1 - \gamma \, \mathrm{cos}(\theta))} \frac{\textbf{x} \otimes \pmb{\gamma}}{\lVert \textbf{x} \rVert} - \gamma \left( \mathrm{cos}(\theta) - \gamma \right) \mathbb{I}\right]
\eeaa

where $\mathrm{cos}(\theta) = (\pmb{\gamma} \cdot \textbf{x})/(\gamma \lVert \textbf{x} \rVert)$

\subsection{Vorton dynamics}
The dynamics of the vortons is given by \cite{Novikov83}:
\begin{eqnarray}
\dot{\x}_\alpha&=&-\frac{1}{4\pi}\sum_{\beta\neq \alpha} \frac{ \rab\times\gamib}{\|\rab\|^3}\\
\dot{\pmb{\gamma}}_{\alpha}&=&-\frac{1}{4\pi}\sum_{\beta\neq \alpha} \bigg[ \frac{\gamia\times\gamib}{\|\rab\|^3}-3\left(\gamia\cdot \rab\right)\frac{(\rab\times \gamib)}{\|\rab\|^5}\bigg].
\label{interactionvorton}
\end{eqnarray}

\end{document}